\documentclass[times,twocolumn,final]{elsarticle}
\usepackage{medima}
\usepackage{framed,multirow}

\usepackage{amsmath,amssymb,amsfonts}
\usepackage{algorithmic}
\usepackage{graphicx}
\usepackage{hyperref}
\usepackage{color}
\usepackage{cleveref}
\usepackage{textcomp}
\usepackage{multirow}
\usepackage{booktabs}

\journal{}

\definecolor{forest}{rgb}{0.16, 0.67, 0.16}

\definecolor{teal}{rgb}{0, 0.5, 0.5}

\definecolor{jade}{rgb}{0, 0.659, 0.42}

\definecolor{indigo}{rgb}{0.294, 0, 0.51}

\newcommand{\win}[1]{\underline{\textbf{#1}}}
\newcommand{\second}[1]{\textbf{#1}}

\newcommand{\signWin}[1]{\textbf{#1}}

\renewcommand{\cite}[1]{\citep{#1}}

\usepackage[normalem]{ulem} 
\newcommand{\stkout}[1]{\ifmmode\text{\sout{\ensuremath{#1}}}\else\sout{#1}\fi}
\newcommand{\citex}[1]{\mbox{\cite{#1}}}

\begin{document}

\verso{Maximilian Springenberg, Annika Frommholz, Markus Wenzel, Eva Weicken, Jackie Ma, and~Nils~Strodthoff}
\begin{frontmatter}
\title{From Modern CNNs to Vision Transformers: Assessing the Performance, Robustness, and Classification Strategies of Deep Learning Models in Histopathology}

\author[1]{Maximilian \snm{Springenberg}}
\author[1]{Annika \snm{Frommholz}}
\author[1]{Markus \snm{Wenzel}}
\author[1]{Eva \snm{Weicken}}
\author[1]{Jackie \snm{Ma}}
\author[2]{Nils \snm{Strodthoff}}

\address[1]{Fraunhofer Heinrich Hertz Institute, Einsteinufer 37, 10587 Berlin, Germany}
\address[2]{University of Oldenburg, Ammerl\"ander Heerstr. 114-118, 26129 Oldenburg, Germany}

\received{TBD}
\finalform{TBD}
\accepted{TBD}
\availableonline{TBD}
\communicated{TBD}

\begin{abstract} 
While machine learning is currently transforming the field of histopathology, the domain lacks a comprehensive evaluation of state-of-the-art models based on essential but complementary quality requirements beyond a mere classification accuracy. In order to fill this gap, we developed a new methodology to extensively evaluate a wide range of classification models, including recent vision transformers, and convolutional neural networks 
such as: ConvNeXt, ResNet (BiT), Inception, ViT and Swin transformer, with and without supervised or self-supervised pretraining. We thoroughly tested the models on five widely used histopathology datasets containing whole slide images of breast, gastric, and colorectal cancer and developed a novel approach using an image-to-image translation model to assess the robustness of a cancer classification model against stain variations. Further, we extended existing interpretability methods to previously unstudied models and systematically reveal insights of the models’ classification strategies that allow for plausibility checks and systematic comparisons. The study resulted in specific model recommendations for practitioners as well as putting forward a general methodology to quantify a model’s quality according to complementary requirements that can be transferred to future model architectures.
\end{abstract}


\end{frontmatter}

\section{Introduction}
\label{sec:introduction}
Machine learning (ML) has the potential to transform the field of histopathology, where expert pathologists visually examine stained tissue specimen under the microscope, e.g., for cancer diagnosis~\cite{spanhol2015dataset}. New ML technologies make the rapid, automatic analysis of large numbers of digitized whole slide images (WSIs) possible and promise to alleviate the burden of the time-consuming examination by human experts in traditional workflows. 
Automatic image analysis also bears the potential to enhance these workflows with quantitative metrics, e.g., by accurately quantifying tumor infiltrating lymphocytes across entire WSIs instead of relying on a coarse visual estimate \cite{klauschen2018scoring}. These innovative improvements are mainly driven by the steady progress of deep learning (DL), which is an ML sub-discipline focusing on multi-layer neural networks. Latest ML methods show very promising results on a broad range of analysis tasks in histopathology (see \cite{Komura2018,Niazi2019,Bera2019,dimitriou2019deep,vanderLaak2021} for recent reviews) and already reach performance levels of human pathologists for specific tasks \cite{Liu2019}.

The predictive performance of an ML model on a given task is typically assessed with standard metrics such as accuracy or sensitivity/specificity. However, additional quality criteria, such as the robustness and interpretability of the model, are highly relevant for the clinical application as well, and are complementary to the quantitative evaluation of the predictive performance. The term robustness has many facets ranging from robustness against shifts in the data distribution, to robustness against input perturbations or adversarial attacks. Concerning interpretability, the field of explainable artificial intelligence (XAI) has witnessed tremendous advances in the past few years and lead to the development of methods that allow certain insights into the decision process of complex ML models (see \cite{lundberg2017unified,JMLR:v22:20-1316,SamPIEEE21,arrieta2020explainable,tjoa2020survey,montavon_2018:methods}~for reviews). XAI techniques have already been adopted in the domain of histopathology to a certain degree (see \cite{Poceviit2020} for a recent review and \cite{hagele_2020:histopatho} for a best-practice paper).

In this paper, we aim to quantitatively compare different state-of-the-art model architectures in the context of histopathology from a comprehensive point of view with respect to predictive performance, interpretability and robustness. On five histopathology datasets, we benchmark convolutional neural networks (CNNs), including recent architectures such as ConvNeXt, but also vision transformers (ViTs) and Swin transformers, where especially ViTs and Swin transformer models have not been explored extensively yet.
Concerning interpretability, we provide relevance (or: attribution) heatmaps for the models with the layer-wise relevance propagation (LRP) framework, and are thereby extending this framework to previously unstudied model architectures. In addition, we propose a quantitative evaluation scheme to assess the plausibility of the resulting relevance heatmaps 
by comparison with tissue components, in particular cell nuclei, which are of particular interest to pathologists in cancer diagnostics as this is where cell division and proliferation processes take place \cite{irshad2013methods,veta2016mitosis,sohail2021multi}.

Finally, we focus on the robustness of the models with respect to stain variations. For a quantitative assessment, we propose to use an image-to-image translation model (CycleGAN) in order to disentangle the effects of stain variation from other sources of distribution shifts. 

Our main contributions are (a) a thorough methodology to compare the performances of a broad range of state-of-the-art image recognition models, including convolutional networks and vision transformers, across five publicly accessible histopathology datasets covering different cancer and tissue types, (b) implementation of LRP for several of the considered models and a novel approach to quantify the focus of attention across the whole dataset using interpretability- paired with segmentation-techniques, and (c) using a new generative adversarial network architecture to measure the robustness of the cancer classification models against staining variations.

\section{Materials and Methods}
\label{sec:materials}

\subsection{Datasets and Tasks}
\label{subsec:datasets}
Five openly accessible datasets serve for the evaluation of the models regarding predictive performance, interpretability, and robustness. The datasets are briefly characterized in \Cref{tab:dataset-overview}.
Most notably, the datasets cover different cancer entities, such as breast cancer (a) metastasis in lymph-nodes (PCam \cite{Veeling2018-qh,EhteshamiBejnordi2017,graham2020dense}) (b) tissue (BreaKHis\cite{spanhol2015dataset}, IDC\cite{Janowczyk2016,CruzRoa2018,CruzRoa2014}), gastric cancer (GasHisSDB(A)\cite{hu2021gashissdb}), and colorectal cancer (MHIST\cite{wei2021petri}). 
Targeted towards identifying invasive ductal carcinoma (IDC) breast cancer tissue, the IDC dataset used in this work replicates the patched IDC dataset from \cite{Janowczyk2016} at a higher spatial resolution using information from the original data sources \cite{CruzRoa2014,CruzRoa2018}.
With a sliding window approach, we join adjacent patches from \cite{Janowczyk2016} creating non-overlapping $100\times100$ patches out of the original $50\times50$ patches filling missing patches by the corresponding parts from the original source \cite{CruzRoa2014,CruzRoa2018}.
We mostly focus on breast cancer throughout this manuscript and use the GasHisSDB (A) and the MHIST datasets to demonstrate that our findings are not specific for breast cancer and can be adapted to other cancer entities. 
Most of the datasets that we included here employ a patch-wise binary classification task, namely ``benign vs. malignant'' (PCam, IDC, GasHis,  MHIST). Only BreaKHis distinguishes different benign and malignant subtypes and is therefore framed as a multi-class classification task.
All datasets had been labeled by expert pathologists.

\begin{table}[h]
\centering
\caption{Overview of the five datasets, with cancer type, number of classes, and sizes of the training/validation/test splits.}
\label{tab:dataset-overview}
\footnotesize
\setlength{\tabcolsep}{2.5pt}
\begin{tabular}{p{1.9cm}p{1.35cm}cllll}
\toprule
{Dataset} & {Cancer type} & {Classes} & {Training} & {Valid.} & {Test} & {Total}\\ \hline
PCam        & Breast (metastasis)   & 2 & 262,144   & 32,768    & 32,768    & 327,680\\
BreaKHis ($\times$40) & Breast    & 8 & 1,005     & 504       & 504       & 2,013\\
IDC         & Breast                & 2 & 26,734    & 10,009    & 16,410    & 53,153\\
GasHisSDB (A)& Gastric               & 2 & 13,313    & 6,657     & 13,314    & 33,284\\
MHIST       & Colorectal            & 2 & 2,175     &  --       & 977       & 3,152\\
\bottomrule
\end{tabular}
\end{table}

\subsection{Models}
\label{subsec:models}
A wide selection of state-of-the-art models with different working principles is included in this study. To guide the architecture selection, we focused mainly on models that scored well on a recent finetuning study \cite{howardOnline}. ResNet (BiT) \cite{he2016deep,Kolesnikov2020}, ConvNeXt \cite{liu2022convnet} and Inception~V3 \cite{szegedy2016rethinking} serve as representatives for modern convolutional neural networks (CNNs). 
Vision transformers (ViTs) are beginning to challenge CNNs in computer vision for natural images, but have so far received only limited attention in the field of histopathology; see \cite{chen2021gashis} for first results in this direction. 
We consider both the original ViT \cite{dosovitskiy2021an} as well as the Swin transformer, a model with a ViT backbone and a more refined handling of patches.
Besides, we also include a BoTNet50 model \cite{srinivas2021bottleneck} where the convolutional layers in the final ResNet-block have been replaced by multi-head self-attention layers, which are the architectural building blocks of transformers. It is worth stressing that even though this latter model architecture is often referred to as a hybrid CNN-Transformer-model, it is more a ResNet model with non-local attention layers rather than an actual transformer model such as the ViT or the Swin transformer.

Recently, efforts have been made to introduce hybrid architectures to histopathology. While the present work focuses on different views on performance as a merit of distinctive architectures, we also feature a hybrid architecture. GasHis \cite{chen2021gashis} features a local and a global feature extraction model, whose feature vectors are concatenated and forwarded to a multilayer perceptron head. With Inception~V3 as local feature model and BotNet-50 as global feature model, we are able to observe whether concatenating architectures into hybrid models has a tangible effect on the performance in the context of histopathology. Complex and light weight variants are featured for ConvNeXt, ViT and Swin architectures with tiny (ConvNeXt-T, ViT-T, Swin-T) and large variants (ConvNeXt-L, Vit-L, Swin-L). To put these models into context, we included two ResNet variants (ResNet-50x1, ResNet-152x2 \citep{Kolesnikov2020}) of similar complexity.

It is common knowledge that large models, in particular those with a low inductive bias such as transformer models \cite{Battaglia2018}, need a comparably large number of samples to develop their full potential, which might distort a comparison where all models are trained from scratch. As nearly all datasets in the field of histopathology have relatively few training samples, we rely on pretraining on ImageNet. We use pretrained weights from a single, common source \citep[the \textit{timm} library;][]{rw2019timm}, in order to ensure comparability in terms of ImageNet pretraining performed according to the latest standards. Where in doubt, we picked the weights pretrained on the largest available dataset: The pretraining datasets are ImageNet for Inception~V3, ImageNet-21k for the ViT and ResNet variants and ImageNet-22k for the Swin transformer and ConvNeXt variants \cite{rw2019timm}.

To gain deeper insight into how much pretraining affects the architectures performance for classification tasks in histopathology, we examined the effects of finetuning with pretrained weights versus training from scratch on PCam. The training procedure of finetuning was kept consistent for all models. AdamW with hyperparameters $\beta_1 = 0.9, \beta_2 = 0.999, \epsilon=10^{-8}$, a weight decay of $\lambda=0.001$ and a learning rate of $0.0005$ was used for 3 epochs on PCam (the fine tuning spans only 15\% of the training from scratch). No benefits were observed from using other optimizers or warm-up strategies, when finetuning. This finetuning procedure was also used for all other datasets, but epochs were increased for smaller datasets.
When training from scratch on PCam, we tailored the training procedures to each architecture individually. All models were trained with a learning rate of $0.0015$. The ResNet variants, as well as the Botnet-50, Inception~V3 and GasHis were trained using AdamW with hyperparameters $\beta_1 = 0.9, \beta_2 = 0.999, \epsilon=10^{-8}$ and a weight decay of $\lambda=0.001$. The transformer variants (ViT, Swin transformer) were trained using stochastic gradient descent with a momentum of $0.9$ and a weight decay of $0.001$. Linear warm-up was used for 4 epochs. Furthermore, no data augmentation was used during warm-up, because we found that training may stall if one applies heavy augmentation when training transformer models from scratch. For the ConvNeXt variants, the same warm-up procedure, as used for the transformer models, was applied before switching to the training configuration of the ResNet variants and other convolutional models using AdamW. In the from scratch training scenario, all models were trained for 20 epochs on PCam.

Model selection was kept consistent for finetuning and training from scratch. The AUC on the validation split was monitored at the end of each epoch and respective models were stored as candidates. Finally, the model with the highest AUC on the validation split was selected for that training run. This choice of model selection is motivated in the AUC being a more reliable measure than accuracy in the presence of smaller datasets. We favour the option to pick from a selection of candidates rather than early stopping, as we found that training can be a bit noisy. The proposed model selection proved to yield good and consistent results. For the MHIST dataset, we experimented with different, custom validation splits to carry out model selection, as the dataset does not provide a validation split in and of itself. However, we found that simply training till convergence yielded much better results on MHIST, than using model selection, for all architectures. This exception could be explained by the very limited amount of data in MHIST. The models in this case seem to benefit more from an extensive training split to calibrate on than from model selection.

With respect to pretraining strategies, recent works \cite{Ciga2022,Srinidhi2022,Azizi2022REMEDIES} demonstrated consistent improvements through self-supervised learning, for example in the form of an intermediate self-supervised training stage starting from ImageNet weights. We deem this as a very promising direction and additionally studied models that had been pretrained by self-supervised contrastive learning specifically on histopathology data, in addition to models pretrained on ImageNet. Recently, advances in applying this technique have been published, such as a pretrained ResNet-50 \citep[`RetCCL',][]{WANG2023102645}, a modified and pretrained Swin-T transformer \citep[`CTransPath',][]{Wang2021, WANG2022102559} and a pretrained ResNet-18 \cite{SelfSupervisedHisto}. Both RetCCL and CTransPath had been trained in a self-supervised setting on large amounts of data, extracted from the cancer genome atlas (TCGA) \citep{Cancer_Genome_Atlas_Research_Network2013-wo} and the pathology AI platform (PAIP) \citep{Kang2021-et} datasets. The self-supervised ResNet-18 was pretrained on a diverse selection of 57 histopathology datasets. We applied the publicly available model weights of these models in two scenarios: (i) frozen feature classification, which is the scenario suggested by the publications cited above, and (ii) initialization for finetuning. When keeping all hidden layers frozen and only tuning the head (i), we observed inferior performance compared to using these weights as an initialization for finetuning the entire model in the same manner, as we did for the pretrained ImageNet weights (ii). Hence, we decided to report only results based on this second approach, identical to the procedure for the ImageNet weights. 

Random hue shifts spanning the entire hue spectrum, horizontal, as well as vertical flips and rotations by up to 180$^\circ$ were included in the augmentation procedure. 
We release the code for training and evaluation of all models in a corresponding code repository \cite{coderepo}.

The models differ in their complexity, which is influenced by the number of parameters 
(ResNet-50x1: 23.5M, ResNet-152x2: 232M, BoTNet50: 18.8M, Inception~V3: 24.3M, ViT-T: 5.5M, ViT-L: 303M, Swin-T: 27.5M, Swin-L: 194M, ConvNeXt-T: 27.8M, ConvNeXt-L: 196M, GasHis: 40.6M), but also by a range of several further factors \citex{hu2021model}, which might play a decisive role when choosing a model in practice.

\subsection{Quality aspects and their evaluation}

\subsubsection{Predictive performance}
\label{subsubsec:predictive-performance}

The performance of the models (see \Cref{subsec:models}) in solving the binary-/multi-classification problems associated with the five datasets (see \Cref{subsec:datasets}, \Cref{tab:dataset-overview}) was assessed with the common performance metrics of accuracy and AUC. As a particular methodological strength of this study, we address two major sources of uncertainty in our assessment.

To address the uncertainty due to the randomness of the training process, we performed $k=5$ training runs for each architecture and report empirical mean as well as standard error over the resulting set of trained models in \Cref{tab:resultspcam}. The statistical uncertainty of the performance results due to the finiteness of the test set was quantified via empirical bootstrapping on the test set (100 iterations). For the statistical comparison of two model architectures, we combine both uncertainties following \cite{Mehari:2021Self}. We claim a set of models $M_1$ to perform not significantly worse than another set of models $M_2$ if in at least $k(k+1) / (2 k^2) = 60\%$ of the $k \times k$ direct comparisons between the runs of both architectures, the 95\% confidence intervals for the respective score differences do overlap with zero or $M_1$ performs better than $M_2$.
The threshold $k(k+1) / (2 k^2)$ is induced by the convention that a set of trained models $M = \{m_i | 1 \leq i \leq k\}$ shall not be significantly worse than $M$ itself, even if the bootstrapping results in an ordering $m_1  \prec m_2 \prec \ldots \prec m_k$, where $m_i \prec m_j$ denotes that $m_i$ performs worse than $m_j$ (i.e. the 95\% confidence intervals obtained via bootstrapping for the respective score differences do not overlap with zero).
Comparisons that resulted in statistically significant differences are indicated with boldface in \Cref{tab:resultspcam}.

\subsubsection{Robustness}

\label{subsec:robustness_methods}
\begin{figure}
\centering
\includegraphics[width=0.95\columnwidth]{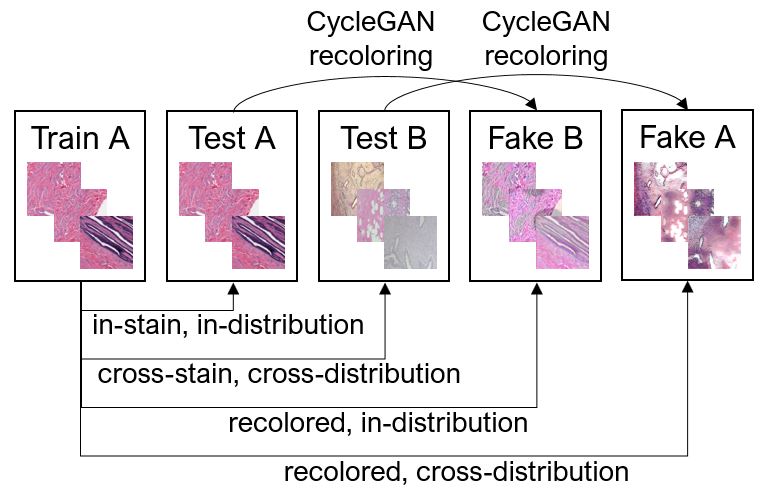}
\caption{Illustration of the robustness evaluation procedure for a model trained on the training split of dataset A (here: BreaKHis or IDC). The assessment consists of four classification tests either on the test split of dataset A (in-stain, in-distribution), or on the test split of dataset B (cross-stain, cross-distribution), on a color-transformed version of the test split of dataset A, referred to as `Fake B' (recolored, in-distribution), or on the color-transformed test split of dataset B 'Fake A' (recolored, cross-distribution).}
\label{fig:cyclegan-illustration}
\end{figure}

Staining of the tissue specimen varies from lab to lab depending on the precise experimental protocol (amount and effective duration of chemical exposure, waiting periods, etc.) and the used scanner equipment \cite{zhao2022RestainNet}, which can result in subtle hue or contrast shifts in the WSIs. Cancer classification models that are planned to be employed in practice should be able to generalize across labs and should be robust against staining variations. We assessed the robustness as follows. First, we trained classification models either on the original BreaKHis or IDC train split. (To allow a direct comparison between both datasets, we trained the binary classifier on BreaKHis using the IDC subclass as positive and all benign samples as negative class.) Then, we evaluated the resulting classifiers separately on the original BreaKHis/IDC test splits and on BreaKHis/IDC images that were recolored with a separate image-to-image translation model. This approach (see \Cref{fig:cyclegan-illustration}) enabled us to to disentangle the effect of color variation from the actual image content of the tissue slices.

Cycle-consistent adversarial networks (CycleGANs) represent a way to train image-to-image translation models based on unpaired data \cite{zhu2017unpaired}. An adaptation of the CycleGAN generator structure by \cite{DEBEL2021102004} has shown to improve the preservation of structural features in stain transformation applications. Similar to fully convolutional encoder-decoder architectures, the image is downsampled to a low-dimensional representation and then upsampled again, trained to retain the relevant features. The generator is an nnU-Net with several skip connections added to learn relevant structural information. Other GAN adaptations that can be used for  \textit{stain normalization} are StainGAN \cite{shaban2018staingan} and RestainNet \cite{zhao2022RestainNet}. All of these implementations aim to perform a template color matching approach in order to adjust the color distribution of an image or dataset to a specified target while retaining image contents.

In our CycleGAN, the generator also has a generic nnU-Net structure \cite{isensee2021nnu}, which is an adaptation of the U-Net specifically for medical image segmentation. The generator output is calculated through a hard \textit{tanh} function as the final activation function resulting in output values between 0 and 1. We hypothesize that the internal connections from the nnU-Net force the model to retain the high-resolution structural components and primarily learn the color transformation.

The discriminator architecture consists of three convolutions followed by a 2D mean pooling layer which reduces the dimensionality of the discriminator output by 1. Each of the convolutions is followed by a 2D batch normalization and an activation function. Firstly and secondly, a \textit{ReLU} activation is used and the third convolution utilizes a \textit{sigmoid} function. The following 2D averaging results in the discriminator relying on color information rather than structural information when comparing original to synthesized images.

We trained 5 CycleGANs for 50 epochs each on the complete BreaKHis dataset (7909 images; incl. the 2,013 images of 40-fold magnification; see \Cref{tab:dataset-overview} and the test set of the IDC dataset (11052 images).
We included the entire BreaKHis dataset of all magnifications for training the CycleGANs, due to the limited number of samples. Concerning the IDC data, the CycleGANs were trained on the IDC test set, because of the similar number of samples compared to the full BreaKHis dataset. Unpaired image-to-image translation works best for a fixed set of images to be recolored if those images are included in the training data. Because the CycleGAN application in this work is limited to the robustness evaluation of the (separate) classifiers, we accept double usage of the IDC test set, aiming for the best recoloring results. Multiple training runs (k=5) for the CycleGAN were implemented to rule out randomness in the generative process.

Data were pre-processed by resizing to a size of $96\times96$ and by random vertical and horizontal flips ($p=0.5$). The learning rate was scheduled to start decaying after half of the training was complete. The generator and discriminator learning rates were set to $0.0002$ and the Adam optimizer ($\beta_1=0.5, \beta_2=0.999$) was used for training. Mean squared error served as loss.
The trained CycleGAN provides two generator models, $G_{\text{BreaKHis} \rightarrow \text{IDC}}$ and $G_{\text{IDC} \rightarrow \text{BreaKHis}}$, for transformations in both directions that enabled us to transform BreaKHis images to mimic the IDC distribution and vice versa. We created images with a \textit{fake IDC} coloring from the original BreaKHis test split, and images with a \textit{fake BreaKHis} coloring from the original IDC test split.

A realistic, meaningful assessment of cross-distribution robustness can only be achieved if characteristics including tissue type, magnification, and image resolution are similar. We adjusted all factors as far as possible for similarity, and used the BreaKHis dataset with an optical magnification of $\times 40$ and the IDC dataset for the robustness validation, because of their similarity in both tissue type and magnification.

\subsubsection{Interpretability}
\label{subsubsec:methods-interpretability}
To gain insights into the opaque decision making of the complex DL models, we draw on the growing body of research literature in the field of XAI, see \cite{montavon_2018:methods,SamXAI19,SamPIEEE21} for reviews. More specifically, we use LRP to calculate relevance heatmaps which highlight regions in input space that speak for or against the classification decision. The LRP-implementations follow \cite{nam2020relative} for ResNet. We implemented LRP rules for the BoTNet and for the Inception model, as well as the ConvNeXt variants along the same lines, none of which have been discussed in the literature so far. This provides a coherent framework for the comparison of heatmaps between different models. To investigate the models' strategies for cancer classification, we propagated the label using LRP for all cancer positive samples of the PCam dataset. As we aimed to identify image regions that are deemed relevant for cancer detection by the model, we averaged over color channels and set all negative relevance to zero, referring to this approach as $R_{\text{mean,max$_0$}}$ pooling. This procedure results in a relevance map emphasizing image regions that are relevant for cancer-positive classification according to LRP.

Next to LRP, we also use attention maps, that are the attention weights of the last layer rearranged into the patch-position of the respective token, as means for XAI.
Bicubic interpolation was used to upscale the 14 $\times$ 14 attention to a 224 $\times$ 224 image.
Unlike LRP, these attention maps carry no direct semantic context regarding classification, but highlight tokens with large attention weights.
Therefore we did not apply any additional pooling to these attention maps.
In this scenario we assume large attention weights to be of similar nature to high, class-positive relevance.
Hence we conducted the same analysis on attention maps, as we did on relevance maps, obtained by LRP.

As one of the main contributions in this work, we put forward a way to quantify relevance maps in terms of the fraction of relevance attributed to certain semantically meaningful subsections of the original image. Here, we distinguish nuclei, background and tissue surrounding nuclei. Nuclei segmentation maps were obtained from an nnU-Net \cite{isensee2021nnu} that was trained on the nuclei segmentation data ``MoNuSeg'' \cite{Kumar2017}. The background was segmented with gray value thresholding and subsequent morphological opening. Pixels that were neither nuclei nor background were considered as tissue surrounding the nuclei.
We then studied the overlap of relevant input areas with the nuclei, tissue, or background segmentation maps. This approach enables us to quantitatively compare different models with respect to their decision making strategies. Potentially, we can find evidence for models using different (or comparable) strategies to achieve a different (or comparable) accuracy. To some degree, the approach also makes it possible to assess the plausibility of the decision making.

Cancer is characterized by rapid cell division activity, which requires first and foremost DNA replication processes in the cell nuclei, and then nuclear division, visible under the microscope. Tumor proliferation rate/speed is therefore an important biomarker for tumor grading and used as a prognostic factor. It is commonly assessed by pathologists counting the number of mitotic nuclei in hematoxylin \& eosin (H\&E) stained histological slides under the microscope and is expressed by the ``mitotic activity index'' (MAI, number of mitoses in 2 mm$^2$ tissue area) \cite{sohail2021multi, veta2016mitosis}. Mitosis describes the process of cell nucleus division across several phases, from the division of the genetic material to the strangulation of the cell body \cite{mcintosh1989mitosis}. For automatic cancer detection, it would therefore be a useful strategy if the ML models were considering cell nuclei in particular.

We accordingly hypothesized that a plausible decision of a model, which classifies tissue as cancerous or not, should accordingly be attributed to a considerable degree to image areas where cell nuclei are located. Otherwise, if the decision of the model can be attributed mostly to other image areas outside of the nuclei or even to the background, we consider this decision as less plausible, and recommend further checks for potential confounding factors that may influence the decision making. 
Overlap between relevance heatmaps and the segmentation maps (of nuclei, tissue, and background) was quantified with the point biserial correlation coefficient \cite{lev1949point,tate1954correlation,kornbrot2014point} (equivalent to Pearson correlation), and with the metric of mass accuracy that is the relevance in the segmentation mask divided by the entire relevance \cite{arras2020ground}.

\section{Results and Discussion}
\label{sec:results}

\begin{table}[htbp]
\centering
\caption{
Performance per dataset. 
Bold and underlined: best model. 
Bold: not significantly worse than best model (see \Cref{subsec:results_predictive_performance}).
}
\label{tab:resultspcam}
\scriptsize
\setlength{\tabcolsep}{3pt}
\begin{tabular}{cll}
\toprule
Model & Acc. [\%] & AUC \\
\hline \hline
\multicolumn{3}{c}{\textit{PCam}}\\
\hline
ResNet-50x1 & 86.73 $\pm$ 0.34 & 0.9544 $\pm$ 0.0026\\
ResNet-152x2& 87.71 $\pm$ 0.46 & 0.9621 $\pm$ 0.0013 \\
BotNet-50 & 87.48 $\pm$ 0.29 & 0.9512 $\pm$ 0.0045 \\
Inception~V3 & 89.40 $\pm$ 0.82 & 0.9671 $\pm$ 0.0031\\
ConvNeXt-T & 88.92 $\pm$ 0.84 & 0.9608 $\pm$ 0.0062 \\
ConvNeXt-L & \win{90.31 $\pm$ 1.27} & \win{0.9722 $\pm$ 0.0045} \\
ViT-T & 85.80 $\pm$ 0.08 & 0.9491 $\pm$ 0.0036 \\
ViT-L & 86.23 $\pm$ 0.54 & 0.9548 $\pm$ 0.0009 \\
Swin-T & 87.78 $\pm$ 0.29 & 0.9576 $\pm$ 0.0032\\
Swin-L & 88.70 $\pm$ 0.58 & 0.9599 $\pm$ 0.0039 \\
GasHis &89.15 $\pm$ 0.84 & 0.9665 $\pm$ 0.0038 \\
CTransPath (Swin-T) & 89.20 $\pm$ 1.07 & 0.9653 $\pm$ 0.0052\\
RetCCL (ResNet-50) & 89.76 $\pm$ 1.47 & 0.9637 $\pm$ 0.0055\\
ResNet-18 (Self-Supervised) & 87.08 $\pm$ 1.36 & 0.9536 $\pm$ 0.0072\\
\hline
\multicolumn{3}{c}{\textit{BreaKHis ($\mathcal{\times}$40)}}\\
\hline
ResNet-50x1 &83.38 $\pm$ 0.95 & 0.9811 $\pm$ 0.0017 \\
ResNet-152x2 &86.72 $\pm$ 0.98 & 0.9853 $\pm$ 0.0014 \\
BotNet-50 & 88.47 $\pm$ 0.25 & 0.9908 $\pm$ 0.0006\\
Inception~V3 & \signWin{92.29 $\pm$ 0.54} & 0.9947 $\pm$ 0.0007\\
ConvNeXt-T & 90.06 $\pm$ 1.23 & 0.9944 $\pm$ 0.0008\\
ConvNeXt-L & 90.02 $\pm$ 0.48 & 0.9926 $\pm$ 0.0016\\
ViT-T & 90.42 $\pm$ 0.66 & 0.9901 $\pm$ 0.0023\\
ViT-L & 86.72 $\pm$ 0.81 & 0.9870 $\pm$ 0.0024\\
Swin-T & 91.13 $\pm$ 0.95 & 0.9937 $\pm$ 0.0014 \\
Swin-L & 92.17 $\pm$ 1.26 & \signWin{0.9940 $\pm$ 0.0006}\\
GasHis & \win{93.00 $\pm$ 1.39} & \win{0.9966 $\pm$ 0.0007}\\
CTransPath (Swin-T) & \signWin{92.25 $\pm$ 0.92} & \signWin{0.9962 $\pm$ 0.0006}\\
RetCCL (ResNet-50) & \signWin{92.33 $\pm$ 0.92} & 0.9944 $\pm$ 0.0007\\
ResNet-18 (Self-Supervised) & \signWin{91.97 $\pm$ 0.51} & \signWin{0.9954 $\pm$ 0.0010}\\
\hline
\multicolumn{3}{c}{\textit{IDC}}\\
\hline
ResNet-50x1 & 87.87 $\pm$ 0.60 & 0.9364 $\pm$ 0.0030\\
ResNet-152x2 &87.54 $\pm$ 1.06 & 0.9377 $\pm$ 0.0023\\
BotNet-50 & 89.07 $\pm$ 0.1 & 0.9440 $\pm$ 0.0008\\
Inception~V3 &89.51 $\pm$ 0.18 & 0.9482 $\pm$ 0.0025 \\
ConvNeXt-T &89.87 $\pm$ 0.35 & 0.9529 $\pm$ 0.0015 \\
ConvNeXt-L & \win{90.12 $\pm$ 0.16} & \win{0.9545 $\pm$ 0.0012}\\
ViT-T & 88.90 $\pm$ 0.37 & 0.9456 $\pm$ 0.0009\\
ViT-L & 88.75 $\pm$ 0.80 & 0.9420 $\pm$ 0.0069\\
Swin-T &89.03 $\pm$ 0.27 & 0.9451 $\pm$ 0.0017\\
Swin-L& 89.88 $\pm$ 0.22 & 0.9529 $\pm$ 0.0016\\
GasHis & 89.24 $\pm$ 0.30 & 0.9479 $\pm$ 0.0008\\
CTransPath (Swin-T) & 89.41 $\pm$ 0.35 & 0.9473 $\pm$ 0.0031\\
RetCCL (ResNet-50) & 89.32 $\pm$ 0.22 & 0.9440 $\pm$ 0.0026\\
ResNet-18 (Self-Supervised) & 88.01 $\pm$ 0.95 & 0.9387 $\pm$ 0.0041\\
\hline
\multicolumn{3}{c}{\textit{GasHisSDB (A)}}\\
\hline
ResNet-50x1 &98.14 $\pm$ 0.09 & 0.9981 $\pm$ 0.0002 \\
ResNet-152x2 & 98.36 $\pm$ 0.12 & 0.9986 $\pm$ 0.0001\\
BotNet-50 & 98.28 $\pm$ 0.07 & 0.9983 $\pm$ 0.0001\\
Inception~V3 & \signWin{99.10 $\pm$ 0.04} & \signWin{0.9995 $\pm$ 0.0001}\\
ConvNeXt-T & \signWin{99.13 $\pm$ 0.06} & \signWin{0.9995 $\pm$ 0.0002}\\
ConvNeXt-L & \win{99.20 $\pm$ 0.04} & \win{0.9996 $\pm$ 0.0001}\\
ViT-T &98.44 $\pm$ 0.09 & 0.9986 $\pm$ 0.0001\\
ViT-L & 98.08 $\pm$ 0.17 & 0.9981 $\pm$ 0.0002\\
Swin-T & \signWin{99.05 $\pm$ 0.04} & \signWin{0.9995 $\pm$ 0.0001} \\
Swin-L & \signWin{98.05 $\pm$ 2.37} & \signWin{0.9986 $\pm$ 0.0021}\\
GasHis & \signWin{99.19 $\pm$ 0.04} & \signWin{0.9995 $\pm$ 0.0001}\\
CTransPath (Swin-T) & 98.31 $\pm$ 0.33 & 0.9989 $\pm$ 0.0002\\
RetCCL (ResNet-50) & 98.77 $\pm$ 0.06 & 0.9992 $\pm$ 0.0000\\
ResNet-18 (Self-Supervised) & 98.83 $\pm$ 0.15 & 0.9993 $\pm$ 0.0002\\
\hline
\multicolumn{3}{c}{\textit{MHIST}}\\
\hline
ResNet-50x1 & \signWin{86.90 $\pm$ 0.50} & 0.9329 $\pm$ 0.0028\\
ResNet-152x2 & \signWin{87.29 $\pm$ 0.78} & 0.9372 $\pm$ 0.0039\\
BotNet-50 & 85.96 $\pm$ 0.85 & 0.9236 $\pm$ 0.0060\\
Inception~V3 & 86.80 $\pm$ 1.19 & 0.9393 $\pm$ 0.0051\\
ConvNeXt-T & \signWin{88.43 $\pm$ 0.34} & \signWin{0.9497 $\pm$ 0.0011}\\
ConvNeXt-L & \win{88.56 $\pm$ 0.67} & \win{0.9539 $\pm$ 0.0009} \\
ViT-T & 84.63 $\pm$ 0.33 & 0.9199 $\pm$ 0.0007\\
ViT-L & 85.81 $\pm$ 1.20 & 0.9235 $\pm$ 0.0097\\
Swin-T & \signWin{86.98 $\pm$ 0.66} & 0.9425 $\pm$ 0.0019\\
Swin-L & \signWin{87.92 $\pm$ 0.27} & \signWin{0.9506 $\pm$ 0.0025}\\
GasHis & \signWin{87.43 $\pm$ 0.56} & \signWin{0.9444 $\pm$ 0.0042}\\
CTransPath (Swin-T) & 83.75 $\pm$ 1.32 & 0.9138 $\pm$ 0.0120\\
RetCCL (ResNet-50) & 84.22 $\pm$ 1.17 & 0.9189 $\pm$ 0.0101\\
ResNet-18 (Self-Supervised) & 81.08 $\pm$ 1.28 & 0.8949 $\pm$ 0.0077\\
\bottomrule
\end{tabular}
\end{table}

\begin{table}[htbp]
\centering
\caption{
Results on PCam, obtained by training from scratch and using pretrained weights (notation as in \Cref{tab:resultspcam}).
}
\label{tab:results-transfer}

\centering
\scriptsize
\setlength{\tabcolsep}{3pt}
\begin{tabular}{cllll}
\toprule
\multirow{2}{*}{Model}   & \multicolumn{2}{c}{Acc. [\%] }& \multicolumn{2}{c}{AUC} \\\cmidrule{2-5}
        & \multicolumn{1}{c}{\textit{pretrained}}   &  \multicolumn{1}{c}{\textit{from scratch}}  &  \multicolumn{1}{c}{\textit{pretrained}}   &  \multicolumn{1}{c}{\textit{from scratch}}\\
\hline 
\hline
\multicolumn{5}{c}{\textit{PCam}}\\
\hline
 ResNet-50x1    & 86.73 $\pm$ 0.34          & 86.91 $\pm$ 0.75  
                & 0.9544 $\pm$ 0.0026       & \win{0.9590 $\pm$ 0.0037}\\
 ResNet-152x2   & 87.71 $\pm$ 0.46          & 85.92 $\pm$ 1.29  
                & 0.9621 $\pm$ 0.0013       & 0.9547 $\pm$ 0.0037\\
 BotNet-50      & 87.48 $\pm$ 0.29          & 80.92 $\pm$ 0.48  
                & 0.9512 $\pm$ 0.0045       & 0.8863 $\pm$ 0.0065\\
 Inception~V3   & 89.40 $\pm$ 0.82          & \win{88.57 $\pm$ 0.48}  
                & 0.9671 $\pm$ 0.0031       & 0.9488 $\pm$ 0.0084\\
 ConvNeXt-T     & 88.92 $\pm$ 0.84          & 78.50 $\pm$ 1.72  
                & 0.9608 $\pm$ 0.0062       & 0.8634 $\pm$ 0.0140\\
 ConvNeXt-L     & \win{90.31 $\pm$ 1.27}    & 78.28 $\pm$ 1.07  
                & \win{0.9722 $\pm$ 0.0045} & 0.8683 $\pm$ 0.0118\\
 ViT-T          & 85.80 $\pm$ 0.08          & 78.75 $\pm$ 0.76  
                & 0.9491 $\pm$ 0.0036       & 0.8795 $\pm$ 0.0080\\
 ViT-L          & 86.23 $\pm$ 0.54          & 78.49 $\pm$ 0.20  
                & 0.9548 $\pm$ 0.0009       & 0.8799 $\pm$ 0.0018\\
 Swin-T         & 87.78 $\pm$ 0.29          & 80.29 $\pm$ 0.39  
                & 0.9576 $\pm$ 0.0032       & 0.8894 $\pm$ 0.0031\\
 Swin-L         & 88.70 $\pm$ 0.58          & 79.93 $\pm$ 0.69  
                & 0.9599 $\pm$ 0.0039       & 0.8911 $\pm$ 0.0057\\   
 GasHis         & 89.15 $\pm$ 0.84          & 88.03 $\pm$ 1.06  
                & 0.9665 $\pm$ 0.0038       & 0.9501 $\pm$ 0.0100\\
\bottomrule
\end{tabular}
\end{table}

\subsection{Predictive performance}
\label{subsec:results_predictive_performance}

The predictive performance of the selected set of models on the five datasets is summarized in \Cref{tab:resultspcam}. 
ConvNeXt-L is the best-performing model with only one exception (GasHis on BreaKHis ($\times$40)) and for both performance metrics (accuracy, AUC).
After the large ConvNeXt-L, its tiny variant ConvNeXt-T, as well as the Inception~V3 and related GasHis hybrid are the next best architectures, when pretraining was applied. 

We additionally trained from scratch on the largest of the featured datasets (PCam). Results are listed in \Cref{tab:results-transfer}. We observe that, without pretraining, all variants of both ViT and Swin transformer show a significant drop in performance. This can be explained by the low inductive bias associated with these models, which rely on large amounts of data \cite{Battaglia2018}. Also the ConvNeXt model shows a significant drop in performance. Training ConvNeXt variants from scratch on PCam proved to be rather unstable and the models struggle in the absence of pretrained weights. 
All of these recently introduced architectures rely on transfer learning to yield good results. Consequently, superseded architectures such as the Inception~V3 or the related GasHis hybrid outperform these models in a from-scratch training scenario. The results, listed in \Cref{tab:results-transfer}, show that all models benefit from pretraining, with the exception of the ResNet-50x1, for which transfer learning showed no significant improvement in performance (the ResNet-50x1 that has been trained from scratch shows a slight, yet statistically significant improvement with respect to both metrics). We advocate for transfer learning on tasks related to histopathology for all complex architectures, especially for transformer- and ConvNeXt-variants. This shows that, especially for models with a low inductive bias, the patterns learned on ImageNet are applicable to non-related classification tasks, as ImageNet is concerned with object classification featuring natural scenes rather than tissue samples.

\begin{figure}[htbp]
    \centering
    \includegraphics[width=0.99\columnwidth]{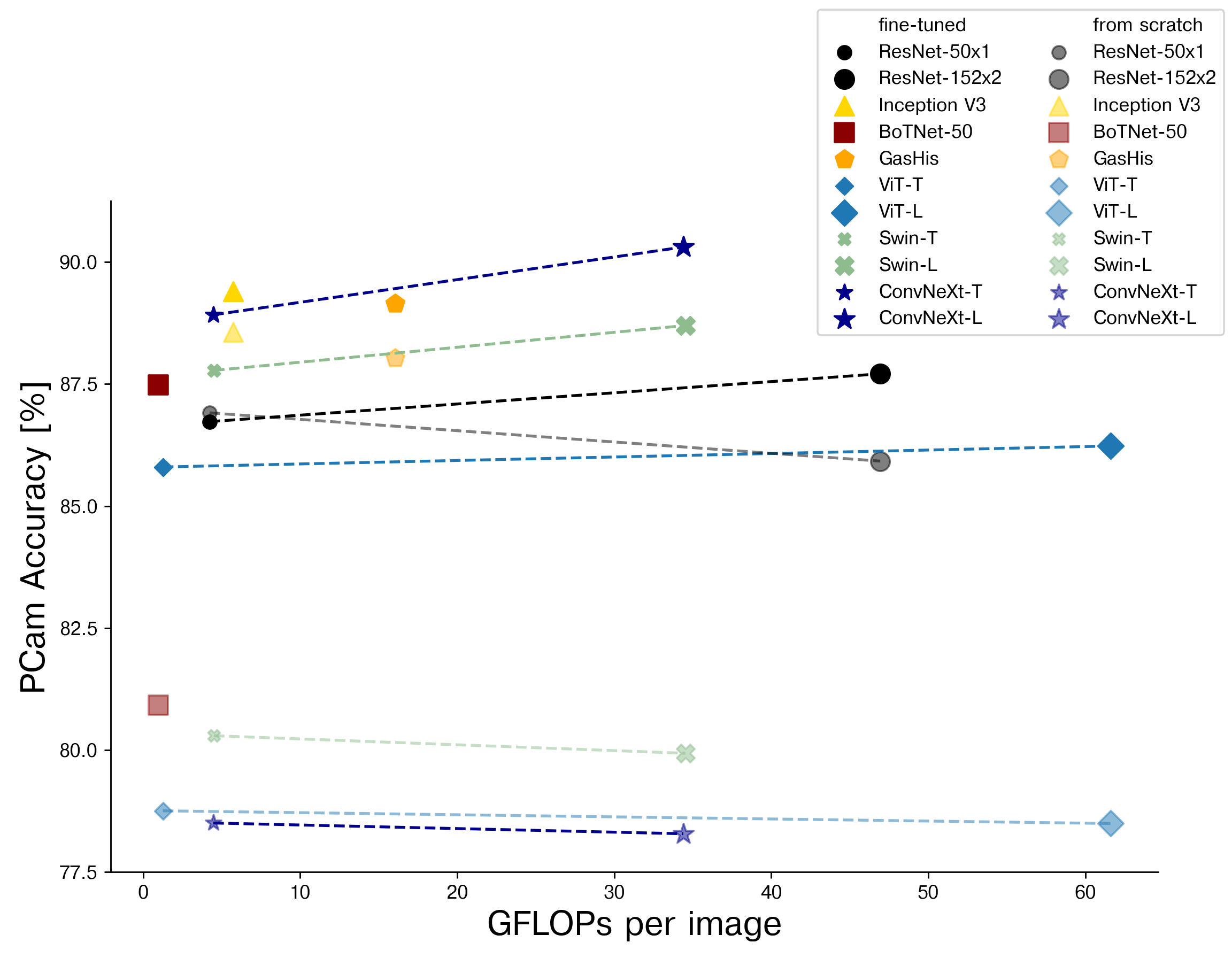}
    \caption{Summary: Accuracy versus model complexity [GFLOPs per image] for fine-tuned models and models trained from scratch on PCam.}
    \label{fig:gflps}
\end{figure}

In \Cref{fig:gflps}, we present a visual summary of our findings for the PCam dataset as the largest and most widely used dataset among the datasets considered in this study. We plot model performance against billion floating point operations (GFLOPs) per image, which is used as a proxy measure for model complexity \citep[calculated with the FLOPs-counter of][]{fvcore2022}. The figure once again demonstrates the positive effect of pretraining. We also observe that competing architectures need a large increase in complexity, with respect to FLOPs per image, to yield significantly better results than the comparably light weight Inception-V3.

The large ConvNeXt-L dominates with respect to both metrics in most cases, followed by GasHis, Inception~V3, the Swin transformer variants and the tiny ConvNeXt-T. The vision transformer consistently scores lower than the Swin transformer, which is coherent with observations on non-histopathology datasets. More complex variants consistently score a little higher than their lightweight counterparts with a few exceptions. On smaller datasets, such as IDC, GasHisSDB (A) and BreaKHis ($\times$ 40), we observe that lightweight variants can outperform more complex configurations of the respective architecture by a tiny margin. This might be an indicator that the complex architectures are more prone to overfitting on small datasets, whereas their lightweight counter parts are apparently better at generalising in this case, due to their limited complexity. However this is not necessarily the case, given that all heavy variants do outperform respective lightweight variants on MHIST, the second smallest dataset, although not always significantly. Regarding our proposed method to highlight performance differences that are statistically significant, we observe that the GasHis model is not significantly worse than the ConvNeXt-L in many cases (BreaKHis ($\times$ 40), GasHisSDB (A), MHIST). 
When GasHis performed not significantly worse than ConvNeXt-L, this was often also the case for Inception~V3, which performed comparably well.
We conclude that GasHis must inherit most of its predictive capacity from the Inception~V3 feature extraction, because BotNet-50 does not perform outstandingly well in comparison to the other architectures. Nevertheless, the additional feature extraction of BotNet-50 apparently allows GasHis to perform marginally better than Inception~V3. While hybrid models have a more diverse feature extraction, we conclude that a superior architecture, such the ConvNeXt-L will outperform hybrids consisting of superseded models in the grand scheme of histopathology.

Turning now to models that had been pretrained by self-supervised contrastive learning specifically on histopathology data, we observe a significant gain in performance for most datasets, with the exception of the small MHIST dataset, where most models struggle to generalise well on the test data and GasHisSDB (A), where performance saturates for all models. While complex models such as ConvNext-L still outperform these self-supervised, pretrained variants, their predictive performance is competitive for most datasets. Compared to the ResNet-50x1, RetCCL benefited greatly from the self-supervised pretraining, with a substantial gain in accuracy on BreaKHis. The performance gain in the case of CTranspath is less pronounced but still significant. The self-supervised ResNet-18 model outperforms the larger ResNet-50x1 model on all datasets, with the sole exception of MHIST, again most notably on BreaKHis. These results provide clear hints for the potential of self-supervised pretraining in histopathology. Nevertheless, the example of BreaKHis compared to MHIST, which have a comparable nominal size, demonstrates that performance gains through self-supervised pretraining cannot be taken for granted but currently still have to be assessed on a case-by-case basis.

To summarize, there are consistent trends in the predictive performance that are stable across both metrics and all datasets, which mostly favor convolutional (ConvNeXt, GasHis, Inception~V3) as opposed to transformer architectures, with the exception of the Swin-L, which is not significantly worse than the best models on three out of five featured datasets.
The best overall architecture ConvNeXt is largely inspired by transformer models, also leveraging the potential of transfer learning. Our analysis suggests using either Inception~V3 as lightweight model (pretrained or even from scratch) or ConvNeXt for an additional performance boost, at a substantially higher level of complexity, for the exploration of new histopathology datasets, notwithstanding the small performance gains that can be achieved by combination with other models into hybrid models.

Furthermore, we have gathered evidence that the presented models are, in principle, well suited for classifying a variety of cancer types, ranging from breast- (PCam, BreaKHis ($\times$ 40), IDC), gastric (GasHisSDB (A)) to colorectal (MHIST) cancer.

\subsection{Robustness}
\label{subsec:results_robustness}

\subsubsection{Recolored test data via CycleGAN}
\label{subsubsec:results_restained}
In an effort to assess the robustness of the models against stain variations, we first generated recolored test images with a CycleGAN (see \Cref{subsec:robustness_methods}) and present the recolored results in this section before turning to the results of the actual robustness tests using these recolored images in the next \Cref{subsubsec:robustness_evaluation}.

\begin{figure}
    \centering
    \includegraphics[width=0.95\columnwidth]{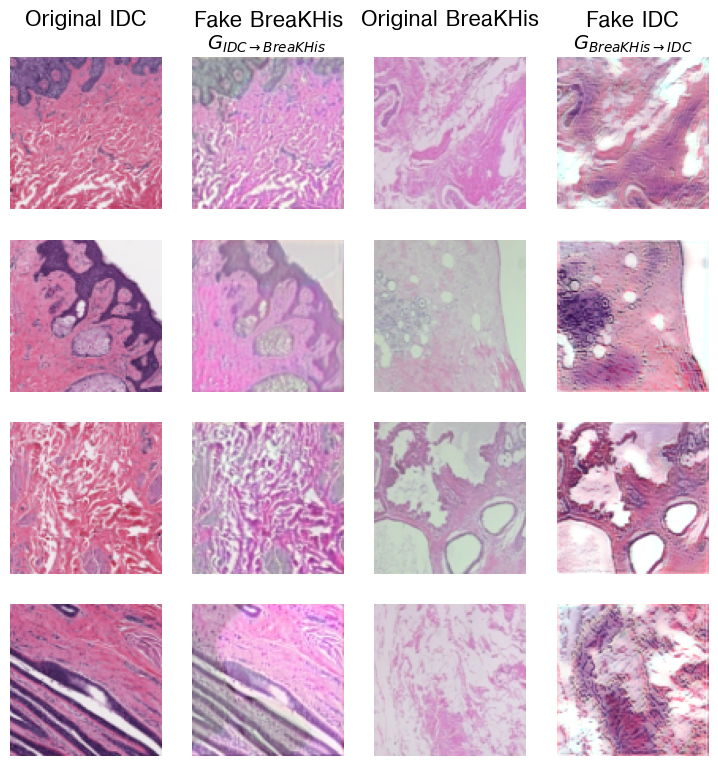}
    \caption{Example images created by the CycleGAN color transformation. Columns 1 and 3 show original images from the IDC and BreaKHis test split. The columns aside are created by the CycleGAN from the original images and imitate the coloring of the respective other dataset.}
    \label{fig:cyclegan-pictures}
\end{figure}

\begin{figure}
    \centering
    \includegraphics[width=\columnwidth]{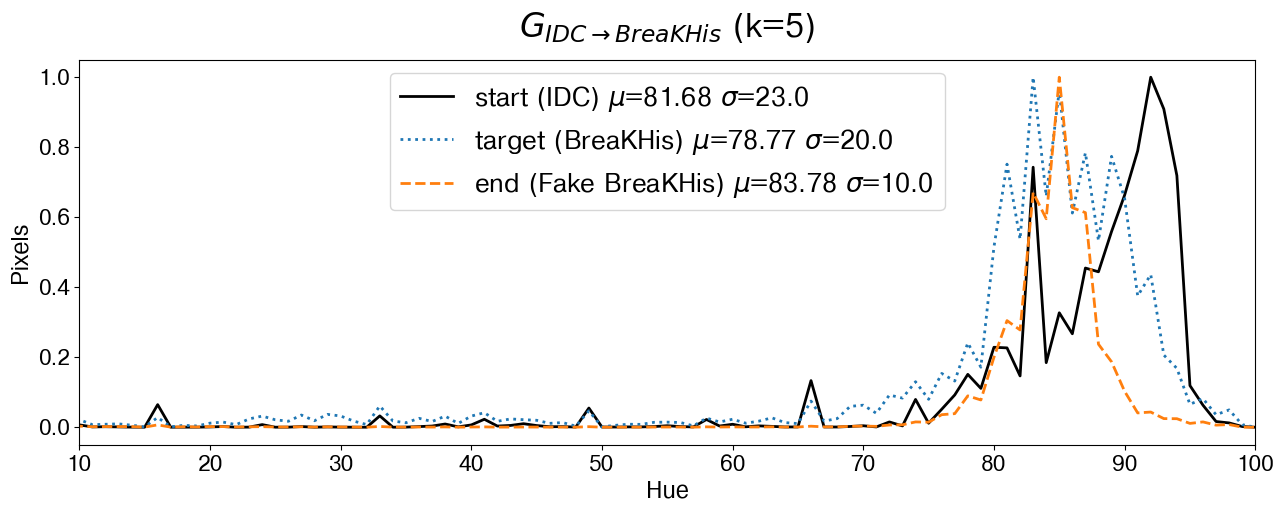}\\
    \includegraphics[width=\columnwidth]{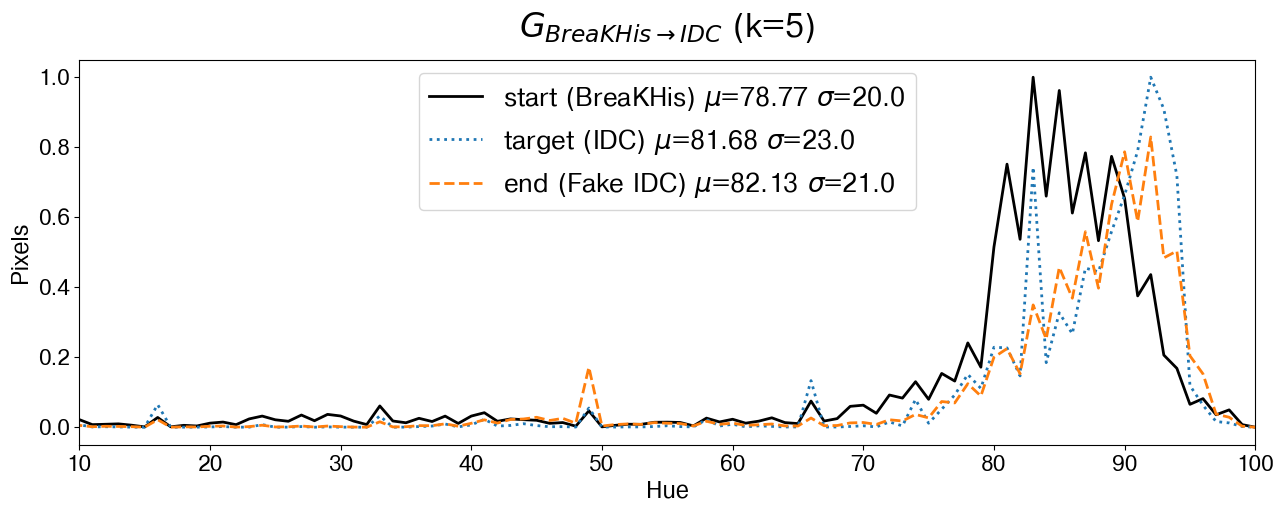} \\
    \caption{
    Hue value histograms for the BreaKHis and IDC datasets and the averaged color transformations in both directions created by a CycleGAN. The blue dotted curve shows the starting hue value distribution for the respective original dataset, the orange dashed curve the resulting hue value distribution at the end of the CycleGAN training 
    (averaged over k=5 training runs) for 50 epochs, and the black solid curve the respective target distribution.
    }
    \label{fig:cyclegan-histograms}
\end{figure}

Exemplary modified versus original images from BreaKHis/IDC are displayed in \Cref{fig:cyclegan-pictures}.
These images stem from the test splits of both datasets and were chosen to represent a variety of original images from the distributions.
To not only rely on a qualitative review of the generated samples, we compared histograms from the CycleGAN-created test sets to the target distribution. 
We chose to display the hue-value of the HSV color space as function values in the histograms to condense the color information into one value for an easier comparison (instead of three values in the case of the RGB color space). 
From the histogram evolution up to 200 epochs, we chose the 
50\textsuperscript{th} epoch as a suitable stopping point for the CycleGAN training. The hue value histograms averaged over k=5 CycleGAN training runs from both transformation directions are shown in \Cref{fig:cyclegan-histograms}. 
The original starting distribution (black) and the target distribution (blue) are added to the diagram, for reference, along with the final distribution after CycleGAN training (orange).

Using the CycleGAN's generator models, $G_{\text{IDC} \rightarrow \text{BreaKHis}}$ and $G_{\text{BreaKHis} \rightarrow \text{IDC}}$, we were able to map the patches' coloring from one datasets distribution to the other and vice versa, such that they mimic another. However, small deviations remain, as can be observed from \Cref{fig:cyclegan-histograms}.
The BreaKHis distribution has two peaks at hue values 83 and 85 and the generators $G_{\text{IDC} \rightarrow \text{BreaKHis}}$ succeeded in displaying the latter peak, but the first peak at 83 is not pronounced as clearly. Generally, the distribution was correctly shifted towards the left but mean value and standard deviation are only achieved to a limited degree. The other generator $G_{\text{BreaKHis} \rightarrow \text{IDC}}$ correctly skewed the BreaKHis dataset to the right but could not produce one distinct peak as seen in the target IDC distribution. This transformation from BreaKHis to IDC succeeded in producing the target mean hue value and standard deviation.

In general, the CycleGANs were able to qualitatively shift the distributions in the correct direction, but the produced fake datasets consistently are narrower in the first case. The reason for this behaviour could be that the generator is able to shift the distribution and skew it based on the shape of the original distribution but struggles with linear histogram stretching. Hence, because the original IDC shape has a narrower peak than the target BreaKHis distribution, the generator is only able to retain the width of the main peak but shifts it to the desired location. 

Furthermore, there is no guarantee that the mapping of color values with the CycleGAN preserves the semantics of the input staining. While the distributions of colors in the staining are approximated to the best of the ability of the CycleGAN, see \Cref{fig:cyclegan-pictures} and \Cref{fig:cyclegan-histograms}, it cannot be taken for granted that the re-coloration retains the staining related to cancerous regions and vice versa. Therefore, the CycleGAN is not to be understood as a stain normalizer, but rather as an agent, that is trying to hand over the classifier-model samples, that are slightly altered in coloration, but not in shape. The shapes of the input have not been altered by the CycleGAN, as we did not allow for such a transformation. Utilising these samples, we can see how much the learned color-bias affects the models performance. Resulting are four tests, yielding information to a models robustness with respect to its distribution- and color biases. (i) A baseline with the samples of the in-distr. test set (in-distr., in-strain.), (ii) testing the color bias with altered images of the in-distr. test set (in-distr., recolored), (iii) testing both color bias and distribution, with samples from a different lab (cross-distr., cross-stain) and (iv) testing both color bias and distribution, with altered images mimicking the in-stain coloring for samples from a different lab (cross-distr., recolored).

Therefore, the CycleGAN is not to be considered as a stain normalizer, for it does not perfectly match the distribution of colors, but rather as a model to mimic a cross-distribution staining.
We consider the generators to produce recolored patches that are more realistic in aesthetic than global transformation methods, such as shifts in hue and saturation, due to their ability to adjust colors locally with respect to visual features.
The generators ability to attain local color transformation can also be observed from the examples displayed in \Cref{fig:cyclegan-pictures}.

\subsubsection{Robustness evaluation}
\label{subsubsec:robustness_evaluation}
In order to single out the specific effect of staining variations between labs, we tested the predictive performance of the classification models on the novel recolored, but in-distribution images (see previous \Cref{subsubsec:results_restained}), and compared the test results with the common setup of train/test-data originating from the same (in-stain/in-distribution) or different sites (cross-stain/cross-distribution).
Moreover, we compared the results with the condition of recolored, cross-distribution test data.
\Cref{tab:robust:BH} lists the results of our robustness evaluation; compare also the illustration in \Cref{fig:cyclegan-illustration}. 
The models consistently perform best on in-stain, in-distribution and better on recolored, in-distribution than on cross-stain, cross-distribution, which matches our expectations.
Regarding the recolored, cross-distribution case, we observe varying results. The fake IDC dataset, obtained from the generator $G_{\text{BreaKHis} \rightarrow \text{IDC}}$ does benefit predictive performance of the models with the exception of the ResNet-152x2 and the BoTNet-50, whereas the fake BreaKHis dataset, obtained from the generator $G_{\text{IDC} \rightarrow \text{BreaKHis}}$ does deteriorate the models performance.
We tie these results to the generators' ability to accurately approximate the respective color distribution, see \Cref{fig:cyclegan-histograms}.
In the following we will discuss the performance drops of the robustness tests (ii), (iii), (iv) in relation to (i) (see previous \Cref{subsubsec:results_restained}).

In the recolored, in-distribution case (ii), models trained on IDC and tested on `Fake BreaKHis' images (that had been recolored with the generator $G_{\text{IDC} \rightarrow \text{BreaKHis}}$) had a drop in performance of about 1.8\% to 5.7\%, whereas models trained on BreaKHis and tested on images with a `Fake IDC' staining (generated with $G_{\text{BreaKHis} \rightarrow \text{IDC}}$) had a drop in performance of about 9.5\% to 18.8\%.
The cross-distribution, cross-staining case (iii) yielded larger drops in performance of 21.2\% to 39.3\% for models trained on IDC, applied on BreaKHis and 23.9\% to 39.2\% for models trained on BreaKHis, applied on IDC.
Finally, in the recolored, cross-distribution case (iv) we observe a drop of 20.0\% to 34.3\% for models trained on IDC and 35.1\% to 46.2\% for models trained on BreaKHis.

Thus, the specific effect of staining variations, or more broadly variations in color, between different labs (ii), which we simulated using a CycleGAN, already does pose a substantial challenge for the cancer classification models examined. 
Nevertheless, the effect of differing distributions between train and test data (iii) appears to be the dominating factor and results in a further large performance decrease (column ``cross-stain, cross-dist.'' in \Cref{tab:robust:BH}). This effect occurs even though we tried to mitigate it as much as possible through weighted sampling of the normal and abnormal classes to match the proportions of the respective original in-distribution test sets.
The effect of recoloring out of distribution test data (iv), using the CycleGAN, does yield slight performance improvements, if the respective generator sufficiently approximates the color distribution of the respective dataset. This can be observed in the recolored, cross-distribution case for models trained on IDC, where all models with the exception of the ResNet-152x2 and the BoTNet-50 perform better than in the cross-distribution, cross-staining case (see Table \ref{tab:robust:BH}, \textit{Trained on IDC}).
The respective generator $G_{\text{BreaKHis} \rightarrow \text{IDC}}$ approximates the color distribution of the IDC dataset well, as can be observed in Figure \ref{fig:cyclegan-histograms}. 
If the generator does not approximate the color distribution sufficiently well, such as the  generator $G_{\text{IDC} \rightarrow \text{BreaKHis}}$ (see Figure \ref{fig:cyclegan-histograms}), we observe an additional drop in performance (see Table \ref{tab:robust:BH}, \textit{Trained on BreaKHis}).

In relation to the baseline (i), significant losses in performance were observed in all robustness tests (ii), (iii) and (iv). We could neither find apparent morphological anomalies in samples misclassified by all models in comparison to the correctly classified samples, nor obvious discrepancies among these samples between the four test conditions (see \Cref{fig:cyclegan-illustration}). To fully understand under which circumstances all models are led to misclassification would, however, require an in-depth morphological investigation on the sample level by histopathologists.

In light of the results, we come to the sobering observation, that it is hard to identify architectures that are particularly robust in either scenario.
No architecture is truly robust to challenges posed by staining variations and distribution shifts.
The results reflect the brittleness of deep learning methods in general. 
Despite the successes in suppressing the proneness of differentiable models to overfitting in recent years, the training dataset still governs the models ability to generalize on out-of-distribution data.
Robustness is not a property that can be taken for granted, but requires dedicated efforts to be achieved and incorporated into the state of the art.

We would like to highlight that, for a sincere evaluation of the classifiers' robustness, we had included global hue value shifts in our data augmentation during training (see \Cref{subsec:models}), inducing a robustness to global color transformations. As a result, classification performance indeed did not significantly change compared to the results from \Cref{tab:resultspcam} when evaluating on test data to which different global color transformations had been applied (data not shown).
The proposed robustness test utilizing the CycleGAN-generated, recolored patches provides more realistic local color transformations and makes it possible to disentangle the cross-distribution- from the stain-variation-effect.

The effects of cross-stain- and cross-distribution-data could be lessened by adapting to distribution shifts in the training data, such as including the respective generators $G_{\text{BreaKHis} \rightarrow \text{IDC}}$, $G_{\text{IDC} \rightarrow \text{BreaKHis}}$ into augmentation or through other domain adaptation techniques \cite{WANG2018135,Guan2021,Kouw2021}, though this is beyond the scope of this work, since we focused on evaluating trained classifiers. 
In particular, our results suggest that the commonly used global color transformations do not lead to enhanced robustness against more realistic stain transformations. 

In summary, we propose to use generative models to assess the robustness against stain variations. Based on our results we cannot single out specific architectures in terms of a particularly high robustness but see a general necessity for further research in this direction.

\begin{table}
    \centering
    \caption{Robustness test results (AUC) for models trained on IDC/BreaKHis and tested on different test data; cf.~\Cref{fig:cyclegan-illustration}. 
    Listed are the mean and standard deviation of five trained models on the respective original test data set, respectively on five \textit{fake} datasets each.
    Bold and underlined: best overall.
    }
    \scriptsize
    \setlength{\tabcolsep}{3pt}
    \begin{tabular}{p{0.18\columnwidth}p{0.18\columnwidth}p{0.18\columnwidth}p{0.18\columnwidth}p{0.18 \columnwidth}}
    \toprule
            Model &
            {in-stain, \vfill in-dist.} &
            {recolored, \vfill in-dist.} &
            {cross-stain, \vfill cross-dist.} &
            {recolored, \vfill cross-dist.} \\ \hline \hline
    \textit{Trained on\vfill IDC} 
    & {\textit{IDC}} & {\textit{Fake BreaKHis}} & {\textit{BreaKHis}} &  {\textit{Fake IDC}}\\
    \hline
        ResNet-50x1     & $0.9364 \pm 0.0030$ & $0.9198 \pm 0.0044$ & $0.5744 \pm 0.0406$ & $0.6746 \pm 0.0408$\\
        ResNet-152x2    & $0.9377 \pm 0.0023$ & $0.9149 \pm 0.0067$ & $0.6548 \pm 0.0590$ & $0.6426 \pm 0.0372$\\
        BoTNet-50       & $0.9440 \pm 0.0008$ & $0.9039 \pm 0.0141$ & $0.7344 \pm 0.0318$ & $0.6523 \pm 0.0358$\\
        Incept.V3       & $0.9482 \pm 0.0025$ & $0.8999 \pm 0.0141$ & $0.7086 \pm 0.0288$ & \win{0.7588 $\pm$ 0.0313}\\
        ViT-T           & $0.9456 \pm 0.0009$ & $0.9246 \pm 0.0060$ & $0.5744 \pm 0.0707$ & $0.6213 \pm 0.0358$\\
        ViT-L           & $0.9420 \pm 0.0069$ & $0.9188 \pm 0.0162$ & $0.6182 \pm 0.1437$ & $0.6758 \pm 0.0435$\\
        Swin-T          & $0.9451 \pm 0.0017$ & $0.9183 \pm 0.0063$ & $0.6111 \pm 0.0798$ & $0.6850 \pm 0.0323$\\
        Swin-L          & $0.9529 \pm 0.0016$ & \win{0.9281 $\pm$ 0.0068} & $0.6300 \pm 0.0479$ & $0.6981 \pm 0.0293$\\
        ConvNeXt-T      & $0.9529 \pm 0.0015$ & $0.9162 \pm 0.0085$ & $0.6770 \pm 0.0319$ & $0.7119 \pm 0.0344$\\
        ConvNeXt-L      & \win{0.9545 $\pm$ 0.0012} & $0.9246 \pm 0.0076$ & $0.6514 \pm 0.0251$ & $0.7413 \pm 0.0329$\\
        {GasHis} & $0.9479 \pm 0.0008$ & $0.8939 \pm 0.0178$ & \win{0.7466 $\pm$ 0.0338} & $0.7518 \pm 0.0329$\\
        \hline
        
    \textit{Trained on\vfill BreaKHis}
    & {\textit{BreaKHis}} & \textit{Fake IDC}& {\textit{IDC}} & { {\textit{Fake BreaKHis}}} \\
    \hline
        ResNet-50x1     & $0.9573 \pm 0.0059$        & $0.8405 \pm 0.0271$ & $0.6431 \pm 0.0308$           & $0.5580 \pm 0.0670$
        \\
        ResNet-152x2    & $0.9604 \pm 0.0029$        & $0.8245 \pm 0.0332$ & $0.6106 \pm 0.0340$           & $0.5163 \pm 0.0727$
        \\
        BoTNet-50       & $0.9720 \pm 0.0045$        & $0.8520 \pm 0.0173$ & $0.6064 \pm 0.0070$           & $0.5993 \pm 0.0633$
        \\
        {Incept.V3}     & $0.9860 \pm 0.0028$        & $0.8868 \pm 0.0199$ & \win{0.7508 $\pm$ 0.0172}     & $0.5891 \pm 0.0599$
        \\
        ViT-T           & $0.9753 \pm 0.0038$        & $0.7917 \pm 0.0393$ & $0.5931 \pm 0.0233$           & $0.5863 \pm 0.0743$
        \\
        ViT-L           & $0.9575 \pm 0.0157$        & $0.8134 \pm 0.0490$ & $0.6383 \pm 0.0309$           & $0.5849 \pm 0.1163$
        \\
        Swin-T          & $0.9681 \pm 0.0195$        & $0.8508 \pm 0.0446$ & $0.6367 \pm 0.0254$           & $0.5976 \pm 0.0715$
        \\
        Swin-L          & $0.9759 \pm 0.0023$        & $0.8830 \pm 0.0240$ & $0.6377 \pm 0.0115$           & $0.6209 \pm 0.0486$
        \\
        ConvNeXt-T      & $0.9599 \pm 0.0046$        & $0.8497 \pm 0.0266$ & $0.6385 \pm 0.0159$           & \win{0.6232 $\pm$ 0.0466}
        \\
        ConvNeXt-L      &$ 0.9649 \pm 0.0053$        & $0.8735 \pm 0.0291$ & $0.6681 \pm 0.0088$           & $0.5707 \pm 0.0396$
        \\
        {GasHis} &\win{0.9866 $\pm$ 0.0042}& \win{0.8900 $\pm$ 0.0214} & $0.7303 \pm 0.0209$           & $0.6082 \pm 0.0568$
        \\
         \bottomrule
    \end{tabular}
    \label{tab:robust:BH}
\end{table}

\subsection{Interpretability}

\label{subsec:results_interpretability}
\subsubsection{Relevant image segments: LRP}
The relevance distributions across nuclei, surrounding tissue, and background, quantitatively characterize the model behaviour (see \Cref{tab:interpretability:LRP}). The classification models appear to attribute relevance mostly to nuclei segments, while surrounding tissue and background seem to be negligible. In other words, nuclei are the most important image segments for the classification decision.

The relevance heatmaps of all models gravitate towards cell nuclei segments with a mass accuracy above the random threshold in every case. 

Pearson correlation between heatmaps and nuclei segments was slightly but significantly ($p<0.05$) positive for GasHis and Inception~V3 and ViT.
On the one hand, this finding aligns very well with the observed strong predictive performance, in the case of GasHis and Inception~V3 (two best-performing models on PCam).
On the other hand, there are also top-performing models such as the 
ConvNeXt variants
that show a comparably low mass accuracy in the nuclei category. 
However, this finding may be related to the LRP relevance maps of ConvNeXt with sharp peaks and slight noise, see \Cref{fig:LRP-maps}.
Obviously, performance does not solely depend on the focus of the model on the most appropriate image regions, but also on the way the information from these regions is combined inside the model, an insight which is not conveyed through relevance heatmaps.

\begin{table}[ht]
    \centering
    \caption{
        Correlation of relevance heatmaps ($R_{\text{mean,max$_0$}}$ pooled) and segmentation maps of nuclei, tissue, background for cancer-positive patches in the PCam test split. 
        Mass accuracy is marked in boldface if it exceeds the score of a random classifier.
       }
    \footnotesize

    \setlength{\tabcolsep}{3pt}
    \begin{tabular}{cccc}
    \toprule
        Model& Mass acc.        & Pearson $r$   & Pearson $p$\\
        \hline\hline
        \multicolumn{4}{c}{\textit{NUCLEI}}\\
        \hline
        {ResNet-50x1}       & \second{0.50 $\pm$ 0.00} & 0.040 $\pm$ 0.002 & 0.072 $\pm$ 0.002\\  
        {ResNet-152x2}      & \second{0.51 $\pm$ 0.00} & 0.050 $\pm$ 0.001 & 0.064 $\pm$ 0.002\\  
        {BoTNet-50}         & \second{0.49 $\pm$ 0.01} & 0.044 $\pm$ 0.010 & 0.061 $\pm$ 0.005\\ 
        {Inception~V3}      & \second{0.51 $\pm$ 0.00} & 0.100 $\pm$ 0.005 & 0.010 $\pm$ 0.000\\ 
        {ConvNeXt-T}        & \second{0.48 $\pm$ 0.01} & 0.004 $\pm$ 0.002 & 0.084 $\pm$ 0.003\\ 
        {ConvNeXt-L}        & \second{0.49 $\pm$ 0.00} & 0.006 $\pm$ 0.001 & 0.080 $\pm$ 0.002\\ 
        {GasHis}            & \second{0.52 $\pm$ 0.00} & 0.093 $\pm$ 0.008 & 0.013 $\pm$ 0.001\\ 
        \hline
        Random              & 0.47        & 0             & $>$0.05\\
        \hline
        \hline
        \multicolumn{4}{c}{\textit{TISSUE}}\\
        \hline
        {ResNet-50x1}       & 0.40 $\pm$ 0.00 & -0.017 $\pm$ 0.001 & 0.084 $\pm$ 0.002\\  
        {ResNet-152x2}      & 0.40 $\pm$ 0.00 & -0.020 $\pm$ 0.002 & 0.076 $\pm$ 0.001\\  
        {BoTNet-50}         & 0.41 $\pm$ 0.00 & -0.019 $\pm$ 0.005 & 0.066 $\pm$ 0.005\\
        {Inception~V3}      & 0.41 $\pm$ 0.00 & -0.033 $\pm$ 0.004 & 0.013 $\pm$  0.001\\ 
        {ConvNeXt-T}        & 0.40 $\pm$ 0.01 & -0.003 $\pm$ 0.002 & 0.090 $\pm$ 0.002\\  
        {ConvNeXt-L}        & 0.38 $\pm$ 0.00 & -0.007 $\pm$ 0.000 & 0.088 $\pm$ 0.002\\ 
        {GasHis}            & 0.40 $\pm$ 0.00 & -0.0423 $\pm$ 0.006 & 0.016 $\pm$ 0.001\\ 
        \hline
        Random & 0.42       & 0             & $>$0.05\\
        \hline
        \hline
        \multicolumn{4}{c}{\textit{BACKGROUND}}\\
        \hline
        {ResNet-50x1}       & 0.09 $\pm$ 0.00 & -0.014 $\pm$ 0.002 & 0.152 $\pm$ 0.003\\  
        {ResNet-152x2}      & 0.09 $\pm$ 0.00 & -0.024 $\pm$ 0.002 & 0.137 $\pm$ 0.003\\  
        {BoTNet-50}         & 0.10 $\pm$ 0.00 & -0.003 $\pm$ 0.011 & 0.109 $\pm$ 0.004\\
        {Inception~V3}      & 0.07 $\pm$ 0.00 & -0.069 $\pm$ 0.006 & 0.025 $\pm$ 0.001\\ 
        {ConvNeXt-T}        & \second{0.11 $\pm$ 0.00} & 0.000 $\pm$ 0.001 & 0.084 $\pm$ 0.003\\  
        {ConvNeXt-L}        & \second{0.12 $\pm$ 0.00} & 0.001 $\pm$ 0.001 & 0.142 $\pm$ 0.001\\ 
        {GasHis}            & 0.09 $\pm$ 0.00 & -0.038 $\pm$ 0.013 & 0.040 $\pm$ 0.001\\   
        \hline
        Random & 0.11              & 0             & $>$0.05\\
    \bottomrule
    \end{tabular}
    \label{tab:interpretability:LRP}
\end{table}

\begin{figure}
    \centering
    \begingroup
    \setlength{\tabcolsep}{1pt} 
    \renewcommand{\arraystretch}{1} 
    \begin{tabular}{c}
        \begin{tabular}{cc}
            \begin{tabular}{cc}
                \rotatebox[origin=c]{90}{\textsf{Input}} &
                \begin{tabular}{c}
                \includegraphics[width=0.3\columnwidth]{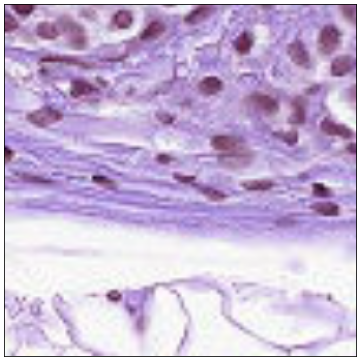}
                \end{tabular}
            \end{tabular} &
            \begin{tabular}{cc}
                \rotatebox[origin=c]{90}{\textsf{Segments}}&
                \begin{tabular}{c}
                \includegraphics[width=0.3\columnwidth]{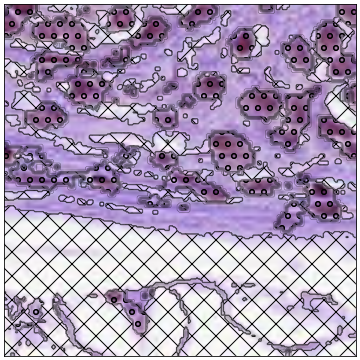}
                \end{tabular}
            \end{tabular} 
        \end{tabular}
        \\
        \begin{tabular}{cc}
            \begin{tabular}{cc}
                \rotatebox[origin=c]{90}{\textsf{ConvNeXt-T}}&
                \begin{tabular}{c}
                \includegraphics[width=0.3\columnwidth]{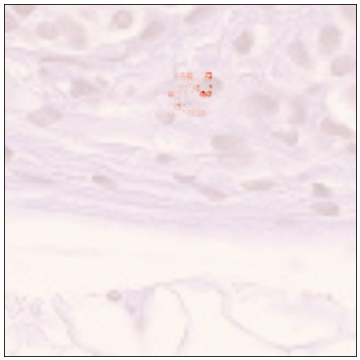} 
                \end{tabular}
            \end{tabular} &
            \begin{tabular}{cc}
                \rotatebox[origin=c]{90}{\textsf{ConvNeXt-L}}&
                \begin{tabular}{c}
                \includegraphics[width=0.3\columnwidth]{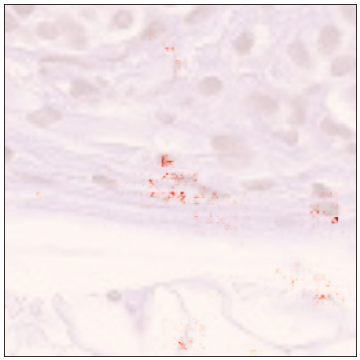}
                \end{tabular}
            \end{tabular}
        \end{tabular}  
        \\
        \begin{tabular}{cc}
            \begin{tabular}{cc}
                \rotatebox[origin=c]{90}{\textsf{ResNet-50x1}}&
                \begin{tabular}{c}
                \includegraphics[width=0.3\columnwidth]{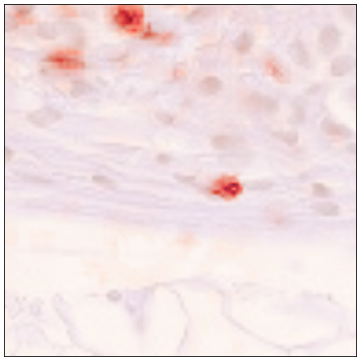} 
                \end{tabular}
            \end{tabular} &
            \begin{tabular}{cc}
                \rotatebox[origin=c]{90}{\textsf{ResNet-152x2}}&
                \begin{tabular}{c}
                \includegraphics[width=0.3\columnwidth]{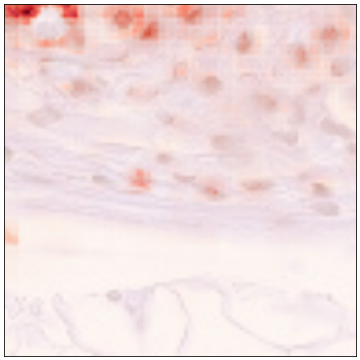}
                \end{tabular}
            \end{tabular}
        \end{tabular} 
        \\
        \begin{tabular}{c c}
            \begin{tabular}{cc}
                \rotatebox[origin=c]{90}{\textsf{BoTNet50}}&
                \begin{tabular}{c}
                \includegraphics[width=0.3\columnwidth]{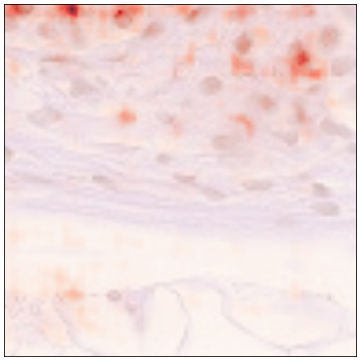} 
                \end{tabular}
            \end{tabular} &
            \begin{tabular}{cc}
                \rotatebox[origin=c]{90}{\textsf{Inception~V3}}&
                \begin{tabular}{c}
                \includegraphics[width=0.3\columnwidth]{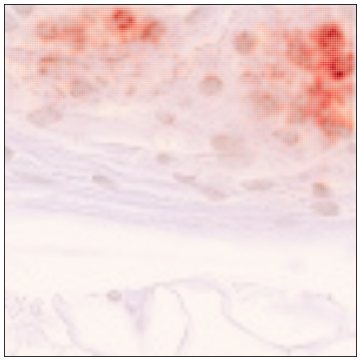} 
                \end{tabular}
            \end{tabular}
        \end{tabular}
        \\
        \begin{tabular}{cc}
            \rotatebox[origin=c]{90}{\textsf{GasHis}}&
            \begin{tabular}{c}
            \includegraphics[width=0.3\columnwidth]{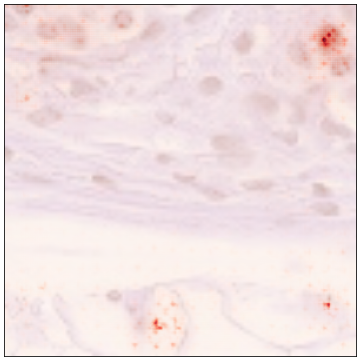} 
            \end{tabular}
        \end{tabular} 
    \end{tabular}
    \endgroup
    \caption{Relevance heatmaps ($R_{\text{mean,max$_0$}}$ pooled) obtained with LRP for different models on an exemplary cancerous (positive label) PCam patch as input.
    The first row shows the input (left) and respective segmentation obtained by the segmentation models (right), dotted-hatching: nuclei; crosshatching: background.
    The following rows display relevance heatmaps obtained by LRP for respective models: ConvNeXt variants, ResNet (BiT) variants, as well as the GasHis and its related architectures BotNet-50 and Inception~V3.
    A correlation with respect to certain image segments can be observed.
    }
    \label{fig:LRP-maps}
\end{figure}

Models show limited statistical correlation according to the Pearson correlation coefficient $r$ and the corresponding $p$-value, however all models mass accuracy shows that all models allocate the most relevance on image segments featuring nuclei. Although there is no clear correlation between predictive performance and relevance allocation to nuclei, we can observe that the nuclei appear to be the most important segments for all models.

It is an interesting observation that even though ConvNeXt-L outperforms all other models by a significant margin, the respective relevance maps do not reflect a stronger focus on nuclei, than relevance maps of other models. This suggests that there are different strategies that can lead to predictive accuracy, based upon inference from individually selected nuclei. While such model-specific strategies can not be explained, we were able to single out the image segments relevant to these strategies.
As a further remark, the focus on nuclei reflects the accordance with strategies used by human pathologists (cf. \Cref{subsubsec:methods-interpretability}). More in-depth interpretability might shed light on alternative classification strategies that are potentially followed by certain ML models. In any case, our analysis provides the first step towards a quantitative evaluation of interpretability methods in the context of histopathology.

It can be observed that the focus on nuclei is not uniformly distributed across the segmented nuclei, e.g., in the examples in \Cref{fig:LRP-maps}.  Relevance is clustered around individual sub-segments, i.e., nuclei, rather than the segmented area of joint nuclei as a whole. Thus, the models distribute high relevance on specific nuclei, while not all nuclei are relevant to the models. This observation aligns with the fact that we cannot distinguish between healthy and cancerous segments of nuclei, considering that the segmentation model was trained on generic nuclei.

In general, all models show very little positive to negative correlation with the segmentation masks of tissue and background. 
GasHis and Inception~V3 show a significant ($p<0.05$) negative Pearson correlation of relevance heatmaps with tissue surrounding the nuclei and, respectively, with the background.
This result confirms the above finding and our hypothesis that strong predictive models indeed focus on cell nuclei and less on surrounding tissue or background, for their classification decision.

\subsubsection{Relevant image segments: attention maps}

Attention-based architectures such as the ViT have the advantage that the architecture-inherent attention maps provide additional insights into the model behavior. Here, we study the attention maps obtained from the classification (CLS)-token of the topmost layer of ViT-L and ViT-T as example.
As the attention heads output and attention maps can vary from training to training, we chose to analyse one model for each variant. We will refer to the $i$-th attention head at the last a layer of a ViT variant, e.g. ViT-L, as ViT-L:$i$ in the following.
Individual ViT heads seem to distribute attention to specific segments and appear to focus mostly on the nuclei and background segments, rather than on tissue (see \Cref{tab:interpretability:Attention}).

\begin{table}[ht]
    \centering
    \caption{
        Correlation of ViT attention (per head) and segmentation maps of nuclei, tissue, background for cancer-positive patches in the PCam test split. Notation as in \Cref{tab:interpretability:LRP}.
    }
    \footnotesize
    \setlength{\tabcolsep}{2.5pt}
    \begin{tabular}{cccccccccc}
    \toprule
        \multirow{2}{*}{Head}   & \multicolumn{3}{c}{\textit{NUCLEI}} 
                                & \multicolumn{3}{c}{\textit{TISSUE}}
                                & \multicolumn{3}{c}{\textit{BACKGROUND}}\\
                        \cmidrule{2-10}
                        & Mass  & $r$   & $p$
                        & Mass  & $r$   & $p$
                        & Mass  & $r$   & $p$\\  
        \hline
        ViT-L:1         & \second{0.58} & 0.019 & 0.023
                        & 0.30          &-0.033 & 0.030
                        & \second{0.12} & 0.008 & 0.062\\
        ViT-L:2         & \second{0.48} & 0.032 & 0.012
                        & \second{0.45} & 0.056 & 0.451
                        & 0.07          &-0.107 & 0.070\\
        ViT-L:3         & 0.45          &-0.026 & 0.012
                        & \win{0.47} & 0.086 & 0.015
                        & 0.08          &-0.083 & 0.021\\
        ViT-L:4         & \win{0.61} & 0.026 & 0.020
                        & 0.21          &-0.042 & 0.024
                        & \second{0.18} & 0.021 & 0.065\\
        ViT-L:5         & 0.45          & 0.010 & 0.014
                        & \second{0.46} & 0.065 & 0.016
                        & 0.9           &-0.084 & 0.021\\
        ViT-L:6         & 0.45          & -0.027 & 0.012
                        & \second{0.46} & 0.077 & 0.014
                        & 0.09          & -0.056 & 0.020\\
        ViT-L:7         & \second{0.50} & 0.083 & 0.009
                        & \second{0.42} & 0.009 & 0.012
                        & 0.07          & -0.117  & 0.019 \\
        ViT-L:8         & 0.44          & -0.039 & 0.011
                        & \second{0.48} & 0.106 & 0.013
                        & 0.08          & -0.078 & 0.020\\
        ViT-L:9         & \second{0.47} & 0.014 & 0.010
                        & \second{0.46} & 0.070 & 0.012
                        & 0.07          & -0.102 & 0.019\\
        ViT-L:10        & \second{0.53} & 0.010 & 0.012
                        & 0.26 & -0.104 & 0.013
                        & \win{0.21} & 0.112 & 0.040\\
        ViT-L:11        & \second{0.49} & 0.051 & 0.012
                        & \second{0.43} & 0.020 & 0.015
                        & 0.08          & -0.097 & 0.019 \\
        ViT-L:12        & \second{0.47} & 0.006 & 0.011
                        & \second{0.46} & 0.073 & 0.013
                        & 0.08          & -0.092 & 0.019\\
        ViT-L:13        & \second{0.50} & 0.066 & 0.011
                        & \second{0.43} & 0.020 & 0.013
                        & 0.07          & -0.097 & 0.019\\
        ViT-L:14        & \second{0.47} & 0.019 & 0.016
                        & \second{0.46} & 0.054 & 0.020
                        & 0.07          & -0.077 & 0.024 \\
        ViT-L:15        & \second{0.54} & 0.015 & 0.024
                        & 0.25          & -0.037 & 0.024
                        & \win{0.21} & 0.037 & 0.066\\
        ViT-L:16        & \second{0.49} & 0.062 & 0.011
                        & \second{0.44} & 0.043 & 0.014
                        & 0.07          & -0.132 & 0.018\\
        \hline
        ViT-T:1         & \second{0.51} & 0.086 & 0.010
                        & \second{0.44} & 0.037 & 0.013
                        & 0.05          & -0.135 & 0.028 \\
        ViT-T:2         & \second{0.52} & 0.133 & 0.008
                        & \second{0.42} & -0.018 & 0.015
                        & 0.06          & -0.142 & 0.020\\
        ViT-T:3         & \second{0.59} & 0.223 & 0.005
                        & 0.38          & -0.088 & 0.008
                        & 0.04          & -0.157 & 0.020\\
                        \hline
                Rand.           & 0.47          & 0     & $>$0.05  
                        & 0.42          & 0     & $>$0.05
                        & 0.11          & 0     & $>$0.05\\
        \bottomrule
    \end{tabular}
    \label{tab:interpretability:Attention}
\end{table}

\begin{figure}
    \centering 
    \small
    \begingroup
    \setlength{\tabcolsep}{1pt}
    \renewcommand{\arraystretch}{1}
    \begin{tabular}{c}
    
        \begin{tabular}{cc}
            \begin{tabular}{cc}
                \rotatebox[origin=c]{90}{\textsf{Input}} &
                \begin{tabular}{c}
                     \includegraphics[width=0.20\columnwidth]{db_plots/revision_interpretability/LRP/original.png}
                \end{tabular}
            \end{tabular} &
            \begin{tabular}{cc}
                \rotatebox[origin=c]{90}{\textsf{Segments}} &
                \begin{tabular}{c}
                    \includegraphics[width=0.20\columnwidth]{db_plots/revision_interpretability/LRP/hatches.png}
                \end{tabular}
            \end{tabular} 
        \end{tabular} 
        \\
        \begin{tabular}{c c c c}
            \begin{tabular}{cc}
                \rotatebox[origin=c]{90}{\textsf{ViT-L:1}} &
                \begin{tabular}{c}
                \includegraphics[width=0.20\columnwidth]{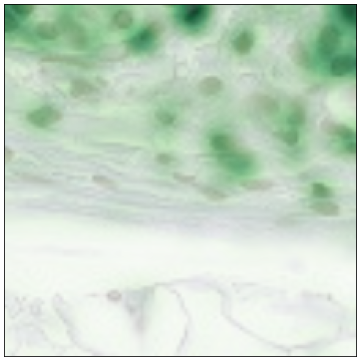} 
                \end{tabular}
            \end{tabular} &
            \begin{tabular}{cc}
                \rotatebox[origin=c]{90}{\textsf{ViT-L:2}} &
                \begin{tabular}{c}
                \includegraphics[width=0.20\columnwidth]{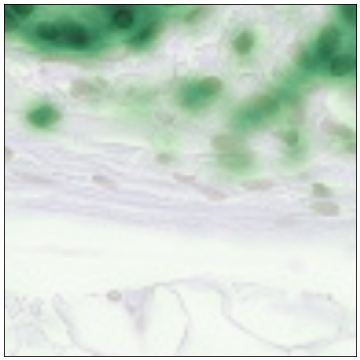} 
                \end{tabular}
            \end{tabular} &
            \begin{tabular}{cc}
                \rotatebox[origin=c]{90}{\textsf{ViT-L:3}} &
                \begin{tabular}{c}
                \includegraphics[width=0.20\columnwidth]{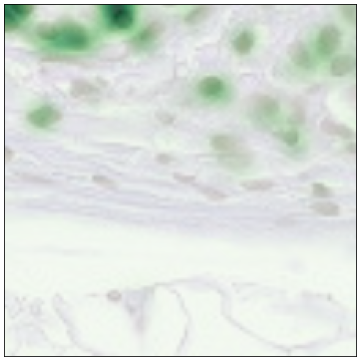} 
                \end{tabular}
            \end{tabular} &
            \begin{tabular}{cc}
                \rotatebox[origin=c]{90}{\textsf{ViT-L:4}} &
                \begin{tabular}{c}
                \includegraphics[width=0.20\columnwidth]{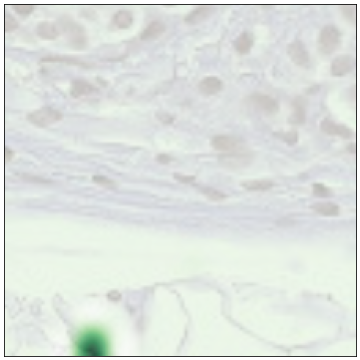}  
                \end{tabular}
            \end{tabular} \\
            \begin{tabular}{cc}
                \rotatebox[origin=c]{90}{\textsf{ViT-L:5}} &
                \begin{tabular}{c}
                \includegraphics[width=0.20\columnwidth]{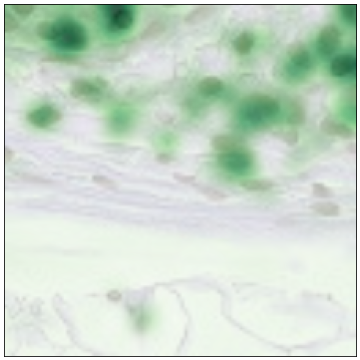} 
                \end{tabular}
            \end{tabular} &
            \begin{tabular}{cc}
                \rotatebox[origin=c]{90}{\textsf{ViT-L:6}} &
                \begin{tabular}{c}
                \includegraphics[width=0.20\columnwidth]{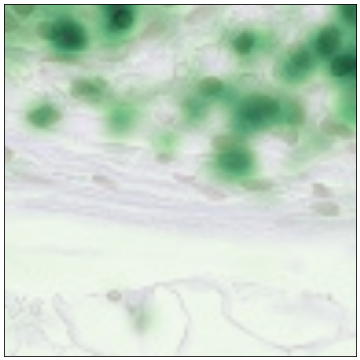}  
                \end{tabular}
            \end{tabular} &
            \begin{tabular}{cc}
                \rotatebox[origin=c]{90}{\textsf{ViT-L:7}} &
                \begin{tabular}{c}
                \includegraphics[width=0.20\columnwidth]{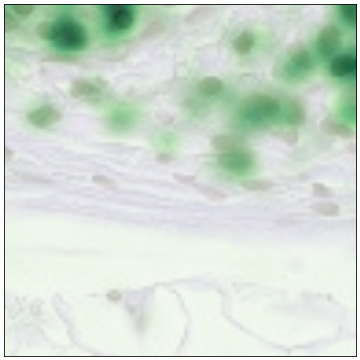} 
                \end{tabular}
            \end{tabular} &
            \begin{tabular}{cc}
                \rotatebox[origin=c]{90}{\textsf{ViT-L:8}} &
                \begin{tabular}{c}
                \includegraphics[width=0.20\columnwidth]{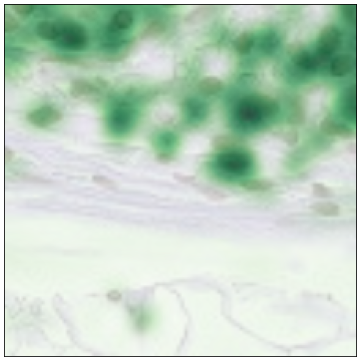} 
                \end{tabular}
            \end{tabular} \\
            \begin{tabular}{cc}
                \rotatebox[origin=c]{90}{\textsf{ViT-L:9}} &
                \begin{tabular}{c}
                \includegraphics[width=0.20\columnwidth]{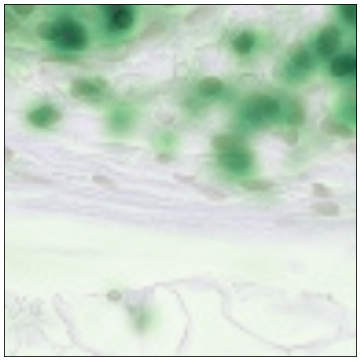} 
                \end{tabular}
            \end{tabular} &
            \begin{tabular}{cc}
                \rotatebox[origin=c]{90}{\textsf{ViT-L:10}} &
                \begin{tabular}{c}
                \includegraphics[width=0.20\columnwidth]{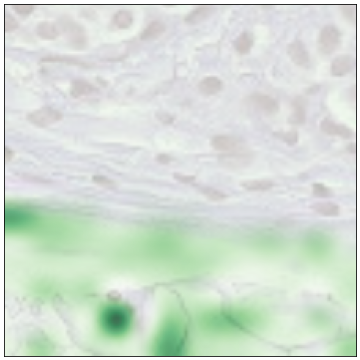} 
                \end{tabular}
            \end{tabular} &
            \begin{tabular}{cc}
                \rotatebox[origin=c]{90}{\textsf{ViT-L:11}} &
                \begin{tabular}{c}
                \includegraphics[width=0.20\columnwidth]{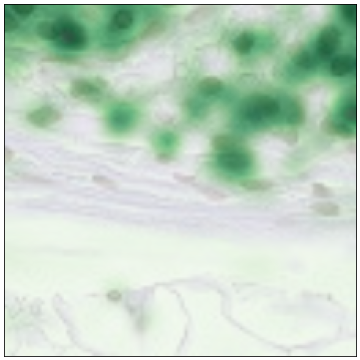} 
                \end{tabular}
            \end{tabular} &
            \begin{tabular}{cc}
                \rotatebox[origin=c]{90}{\textsf{ViT-L:12}} &
                \begin{tabular}{c}
                \includegraphics[width=0.20\columnwidth]{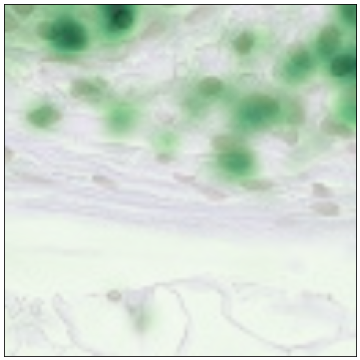} 
                \end{tabular}
            \end{tabular} \\
            \begin{tabular}{cc}
                \rotatebox[origin=c]{90}{\textsf{ViT-L:13}} &
                \begin{tabular}{c}
                \includegraphics[width=0.20\columnwidth]{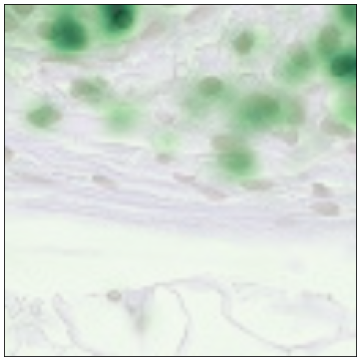} 
                \end{tabular}
            \end{tabular} &
            \begin{tabular}{cc}
                \rotatebox[origin=c]{90}{\textsf{ViT-L:14}} &
                \begin{tabular}{c}
                \includegraphics[width=0.20\columnwidth]{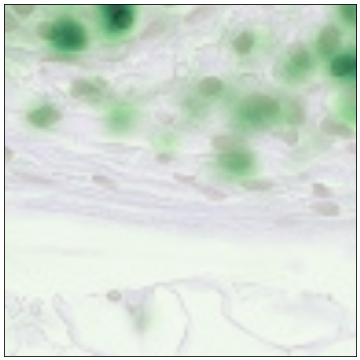} 
                \end{tabular}
            \end{tabular} &
            \begin{tabular}{cc}
                \rotatebox[origin=c]{90}{\textsf{ViT-L:15}} &
                \begin{tabular}{c}
                \includegraphics[width=0.20\columnwidth]{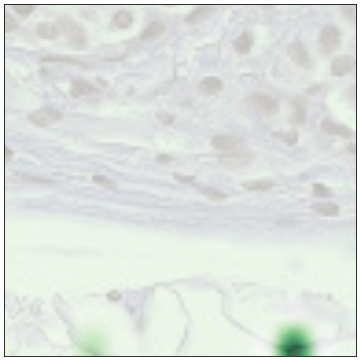} 
                \end{tabular}
            \end{tabular} &
            \begin{tabular}{cc}
                \rotatebox[origin=c]{90}{\textsf{ViT-L:16}} &
                \begin{tabular}{c}
                \includegraphics[width=0.20\columnwidth]{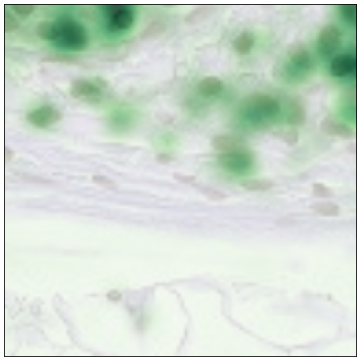} 
                \end{tabular}
            \end{tabular}
        \end{tabular}
        \\
        \begin{tabular}{c c c}
            \begin{tabular}{cc}
                \rotatebox[origin=c]{90}{\textsf{ViT-T:1}} &
                \begin{tabular}{c}
                \includegraphics[width=0.20\columnwidth]{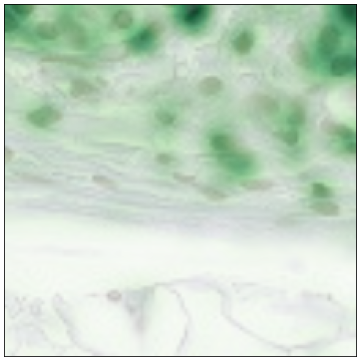} 
                \end{tabular}
            \end{tabular} &
            \begin{tabular}{cc}
                \rotatebox[origin=c]{90}{\textsf{ViT-T:2}} &
                \begin{tabular}{c}
                \includegraphics[width=0.20\columnwidth]{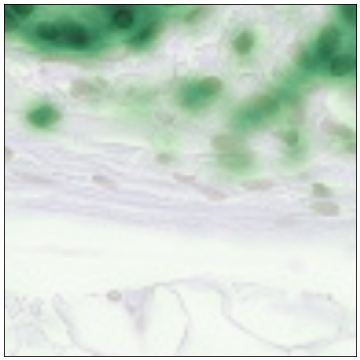} 
                \end{tabular}
            \end{tabular} &
            \begin{tabular}{cc}
                \rotatebox[origin=c]{90}{\textsf{ViT-T:3}} &
                \begin{tabular}{c}
                \includegraphics[width=0.20\columnwidth]{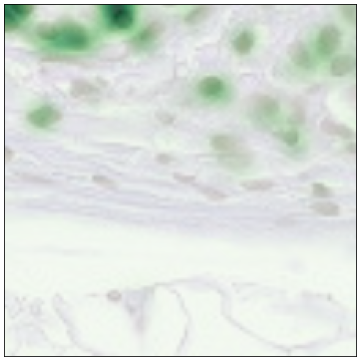} 
                \end{tabular}
            \end{tabular} 
        \end{tabular}
    \end{tabular}
    \endgroup
    \caption{Attention maps (green) per head in the last ViT layer given a PCam patch.
    Dotted-hatching: nuclei; crosshatching: background.
    The 14$\times$14 attention maps were upscaled with bilinear interpolation.
    }
    \label{fig:attn-maps}
\end{figure}

Due to the ability of the ViT's final layer to join attention maps, the model may even divide its attention into multiple heads. Furthermore, a head does not have to focus on one thematic segment, that is to say it may deal with certain aspects of different segment classes at once, which can be observed in \Cref{fig:attn-maps} and \Cref{tab:interpretability:Attention}. Our methodology provides a quantitative way to assess earlier qualitative findings on the specialization of attention heads in the literature \cite{chen2022self}.
While all segments are relevant to at least one head of the ViT-L, the main focus of the model's heads appear to be on the nuclei, much like convolutional architectures. As the ViT-T features only three attention heads, we see that all heads mostly focus on nuclei, with some attention bleeding into surrounding tissue, see \Cref{fig:attn-maps}.

To summarize, interpretability methods make it possible to gain insights into the strategies pursued by different models to reach the classification decision. Our main contribution in this respect is correlating the corresponding heatmaps with the output of segmentation models to assess quantitatively and beyond single examples how much relevance the models put on different image compartments. Whereas all models put most relevance on nuclei, there are hints for distinct model strategies that can lead to a comparable predictive performance.

\section{Conclusion}

Relating back to the main achievements brought forward in the introduction, we summarize the three main findings of our study as follows. 
(a) In a thorough benchmarking study involving a broad range of state-of-the art image recognition models across five public histopathology datasets, we identify the lightweight Inception V3 (pretrained or even from scratch) and the considerably more complex, pretrained ConvNeXt-L, for an additional performance boost,
as best-performing models. 
(b) Our quantitative assessment of attribution (heat)maps reveals that all models focus on nuclei. Slight differences in the relevance distribution can be seen as an indication for different classification strategies, which can be exploited, e.g., through the formation of hybrid/ensemble models. (c) Our proposed methodology allows to quantitatively assess the robustness against staining differences and revealed, irrespective of the model architecture, insufficient robustness (see \Cref{subsubsec:robustness_evaluation}) as a major obstacle for the application of deep-learning-based image recognition models in the wild.

\section*{Acknowledgment}
The authors would like to thank Angel Cruz-Roa and Anant Madabhushi for helpful correspondence on the IDC dataset and the anonymous reviewers for their valuable comments.
This work was partly funded by the Fraunhofer Society, by the German Federal Ministry for Education and Research as Patho234 (ref.\ 031LO207) and BIFOLD -- Berlin Institute for the Foundations of Learning and Data (ref.\ 01IS18025A and ref.\ 01IS18037A).

\bibliographystyle{model2-names.bst}\biboptions{authoryear}
\bibliography{bibfile.bib}

\end{document}